\documentclass[%
nofootinbib,
prc,
reprint,
superscriptaddress,
showpacs,preprintnumbers,
 amsmath,amssymb,
 aps,
]{revtex4-1}

\usepackage{comment}
\usepackage{graphicx}
\usepackage{dcolumn}
\usepackage{bm}
\usepackage{enumitem}
\usepackage{color}
\usepackage{multirow}
\usepackage{lineno,hyperref}

\newcommand{\fig}[1]{fig.~\ref{#1}}
\newcommand{\tab}[1]{table~\ref{#1}}
\newcommand{\Sec}[1]{sec.~\ref{#1}}
\newcommand{\eq}[1]{eq.~(\ref{#1})}

\newcommand{\al}{$\alpha$}
\newcommand{\g}{$\gamma$}
\newcommand{\raa}{($\alpha$,$\alpha$)}
\newcommand{\raX}{($\alpha$,$X$)}

\newcommand{\rag}{($\alpha$,$\gamma$)}
\newcommand{\ran}{($\alpha$,n)}

\newcommand{\rann}{($\alpha$,2n)}

\newcommand{\rap}{($\alpha$,p)}

\newcommand{\rga}{($\gamma$,$\alpha$)}

\newcommand{\stot}{$\sigma_{\rm{reac}}$}

\newcommand{\auvii}{$^{197}$Au}
\newcommand{\tlix}{$^{199}$Tl}
\newcommand{\tlnull}{$^{200}$Tl}
\newcommand{\tli}{$^{201}$Tl}

\newcommand{\Nsv}{$N_A$$\left< \sigma v \right>$}
\newcommand{\fdev}{$f_{\rm{dev}}$}
\newcommand{\fdevbar}{$\bar{f}_{\rm{dev}}$}

\begin{document}

\preprint{}	

\title{Cross section of $\alpha$-induced reactions on $^{197}$Au at sub-Coulomb energies}

\author{T.~Sz\"ucs}%
\email{szucs.tamas@atomki.mta.hu}%
\affiliation{Institute for Nuclear Research (MTA Atomki), H--4001 Debrecen, Hungary}%
\affiliation{Helmholtz-Zentrum Dresden-Rossendorf (HZDR), D--01328 Dresden, Germany}

\author{P.~Mohr}%
\affiliation{Institute for Nuclear Research (MTA Atomki), H--4001 Debrecen, Hungary}
\affiliation{Diakonie-Klinikum, D--74523 Schwäbisch Hall, Germany}

\author{Gy.~Gy\"urky}%
\affiliation{Institute for Nuclear Research (MTA Atomki), H--4001 Debrecen, Hungary}

\author{Z.~Hal\'asz}%
\affiliation{Institute for Nuclear Research (MTA Atomki), H--4001 Debrecen, Hungary}

\author{R.~Husz\'ank}%
\affiliation{Institute for Nuclear Research (MTA Atomki), H--4001 Debrecen, Hungary}

\author{G.~G.~Kiss}%
\affiliation{Institute for Nuclear Research (MTA Atomki), H--4001 Debrecen, Hungary}

\author{T.~N.~Szegedi}%
\affiliation{Institute for Nuclear Research (MTA Atomki), H--4001 Debrecen, Hungary}
\affiliation{University of Debrecen, H--4001 Debrecen, Hungary}

\author{Zs.~T\"or\"ok}%
\affiliation{Institute for Nuclear Research (MTA Atomki), H--4001 Debrecen, Hungary}

\author{Zs.~F\"ul\"op}%
\affiliation{Institute for Nuclear Research (MTA Atomki), H--4001 Debrecen, Hungary}

\date{\today}

\begin{abstract}

\begin{description}
\item[Background]
  Statistical model calculations have to be used for the determination of
  reaction rates in large-scale reaction networks for heavy-element
  nucleosynthesis. A basic ingredient of such a calculation is the
  $\alpha$-nucleus optical model potential. Several different parameter sets
  are available in literature, but their predictions of $\alpha$-induced
  reaction rates vary widely, sometimes even exceeding one order of
  magnitude.
\item[Purpose]
  This paper presents the result of $\alpha$-induced reaction cross-section
  measurements on gold which could be carried out for the first time very
  close to the astrophysically relevant energy region. The new experimental
  data are used to test statistical model predictions and to constrain the
  $\alpha$-nucleus optical model potential. 
\item[Method] For the measurements the activation technique was used. The cross section of the ($\alpha$,n) and ($\alpha$,2n) reactions was determined from $\gamma$-ray counting, while that of the radiative capture was determined via X-ray counting.
\item[Results]
The cross section of the reactions was measured below E$_{\alpha} = 20.0$~MeV. In the case of the $^{197}$Au($\alpha$,2n)$^{199}$Tl reaction down to 17.5~MeV with 0.5-MeV steps, reaching closer to the reaction threshold than ever before.
The cross section of $^{197}$Au($\alpha$,n)$^{200}$Tl and $^{197}$Au($\alpha$,$\gamma$)$^{201}$Tl was measured down to E$_{\alpha} = 13.6$ and 14.0~MeV, respectively, with 0.5-MeV steps above the ($\alpha$,2n) reaction threshold and with 1.0-MeV steps below that. 
\item[Conclusions]
  The new dataset is in agreement with the available values from the literature, but is more precise and extends towards lower energies. Two orders of magnitude lower cross sections were successfully measured than in previous experiments which used $\gamma$-ray counting only, thus providing experimental data at lower energies than ever before. The new precision dataset allows us to find the best-fit $\alpha$-nucleus optical model potential and to predict cross sections in the Gamow window with smaller uncertainties.
\end{description}
\end{abstract}

\maketitle

\section{Introduction}
The elements heavier than iron are mainly produced via neutron capture reactions \cite{B2FH57, Kappeler11-RMP, Arnould07-PR}. These processes, however, cannot create the so-called $p$-nuclei on the proton-rich side of the
valley of stability. The so-called $\gamma$ process \cite{Rauscher13-RPP} is mainly responsible for the production of these isotopes. The $\gamma$ process occurs in hot, dense astrophysical plasma environments like in thermonuclear
supernovae \cite{Nishimura18-MNRAS,Travaglio18-AJ} or in core collapse supernovae \cite{Woosley78-AJS,Rauscher02-AJ}. The $\gamma$-process reaction network
involves tens of thousands of reactions on thousands of mainly unstable nuclei, thus the reaction rates have to be predicted in a wide mass and temperature range. For this purpose the Hauser-Feshbach (H-F) statistical model \cite{Hauser52-PR} using global optical model potentials (OMPs) is widely used.  While the nucleon-nucleus OMP (N-OMP) is relatively well known, the predicted reaction rates may vary over one order of magnitude depending on the chosen $\alpha$-nucleus OMP (A-OMP) \cite{Kiss13-PRC,Netterdon13-NPA}. Recently, several cross-section measurements have been carried out mostly on proton rich isotopes to test the global A-OMPs
(e.g., \cite{Ozkan07-PRC,Sauerwein11-PRC,Filipescu11-PRC,Netterdon13-NPA,Kiss14-PLB,Simon15-PRC,Yalcin15-PRC,Halasz16-PRC,Scholz16-PLB,Mayer16-PRC,Szucs18-PLB,Korkulu18-PRC,Kiss18-PRC}).

In the present work the cross sections of the \auvii \rag \tli , \auvii \ran \tlnull , and \auvii \rann \tlix\ reactions were measured at energies below the Coulomb barrier, reaching the upper end of the Gamow window for typical temperatures of the $\gamma$ process of $T_9 \approx 2 - 3$ (where $T_9$ is the temperature in $10^9$ K). The new experimental results are compared to the predictions of several open-access global A-OMPs. Although \auvii\ is not in the $p$-nuclei mass range, it is only slightly above the heaviest $p$-nucleus $^{196}$Hg, thus it can help understanding the systematic in the mass region. Furthermore, experimental studies are
facilitated by the mechanical and chemical properties of gold and by the fact
that gold is mono-isotopic with the only stable isotope \auvii . The
application of the activation technique \cite{Gyurky19-EPJA} is possible because of the reasonable
half-lives of the residual \tlix , \tlnull , and \tli\ nuclei. However,
$\gamma$-ray spectroscopy had to be complemented by X-ray spectroscopy to
cover all reactions under study in the present work, and the X-ray decay
curves had to followed for a long period to disentangle the contributions of the
different reaction channels.

The paper is organized as follows. In \Sec{sec:react} the reactions under investigation will be presented. In \Sec{sec:exp} the experimental details will be given, while in \Sec{sec:analy} the data analysis is detailed. The experimental results are summarized in \Sec{sec:res}. In \Sec{sec:theo} the obtained data are compared to statistical model calculation. Finally in \Sec{sec:sum} a summary is given.
\begin{table*}[t]
\caption{Decay parameters of the reaction products of $\alpha$-induced reactions on $^{197}$Au \cite{NDS199,NDS200,NDS201}.}
\label{tab:param}
\center
\begin{ruledtabular}
\begin{tabular}{l l c c c c c}
\multirow{2}{*}{Reaction}	&Reaction	& Half-life	& X- or $\gamma$-ray& $I_{rel_i}$ : Relative		& $M$ : Multiplicator		& Absolute \\
							& product		& (h)	& energy (keV)		& intensity (\%)	& 	for absolute intensity				& intensity (\%) \\
\noalign{\smallskip}\colrule\noalign{\smallskip}			
($\alpha$,$\gamma$)	&	$^{201}$Tl	& 73.01\,$\pm$\,0.04& 70.8	& 					& 					& 44.6\,$\pm$\,0.6\\
				&				&					& 167.4			& 					&					& 10.00\,$\pm$\,0.06\\
\noalign{\smallskip}\colrule\noalign{\smallskip}		
($\alpha$,n)	&	$^{200}$Tl	& 26.1\,$\pm$\,0.1		& 367.9			& 100				& \multirow{4}{*}{0.87\,$\pm$\,0.06}	& 87\,$\pm$\,6\\
				&				&					& 579.3			& 15.8\,$\pm$\,0.8	& 					& 13.7\,$\pm$\,1.2\\
				&				&					& 828.3			& 12.4\,$\pm$\,0.7	& 					& 10.8\,$\pm$\,1.0\\
				&				&					& 1205.6		& 34.4\,$\pm$\,1.9	& 					& 30\,$\pm$\,3\\			
\noalign{\smallskip}\colrule\noalign{\smallskip}	
($\alpha$,2n)	&	$^{199}$Tl	& 7.42\,$\pm$\,0.08	& 158.4			& 40\,$\pm$\,2		& \multirow{4}{*}{0.124\,$\pm$\,0.012}& 5.0\,$\pm$\,0.5\\
				&				&					& 208.2			& 99\,$\pm$\,5		& 	 				& 12.3\,$\pm$\,1.3\\
				&				&					& 247.3			& 75\,$\pm$\,4		&					& 9.3\,$\pm$\,1.0\\
				&				&					& 455.5			& 100\,$\pm$\,5		&					& 12.4\,$\pm$\,1.4\\
\end{tabular}
\end{ruledtabular}
\end{table*}

\section{\label{sec:react}Studied reactions}
Gold has only one stable isotope the  $^{197}$Au, therefore the 100\% isotopic purity of the targets is naturally granted. In the energy range investigated here not only radiative $\alpha$ capture occurs, but capture can also be followed by  one or two neutron emissions\footnote{Many more reaction channels - involving mostly $\alpha$ emissions - are energetically possible, but they typically have much lower cross sections - because of the Coulomb barrier in the exit channel - than the ones studied in the present work.}. All these three reaction channels - detailed below - lead to radioactive nuclei, thus the activation technique can be used to investigate them. 

In \tab{tab:param} the decay parameters of the reaction products are summarized.
In the case of the $\gamma$ rays the absolute intensities are obtained from the relative intensities and a multiplicator as given in the decay data compilations \cite{NDS199,NDS200,NDS201}. These are indicated in the table and used in the analysis in order to take the systematic uncertainties correctly into account. In the case of the X ray the absolute intensities per decay is the available quantity in the compilation \cite{NDS201}.

\subsection{$^{197}$Au($\alpha$,$\gamma$)$^{201}$Tl}
There is only one cross-section dataset for this reaction in the literature by Basunia \textit{et al.} \cite{Basunia07-PRC} and several derived upper limits \cite{Capurro88-JRNC, Necheva97-ARI, Kulko07-PAN}. Even if Capurro \textit{et al.} \cite{Capurro88-JRNC} published their data as definite values, they can only be considered as upper limits because of some neglected experimental issues as pointed out by Necheva \textit{et al.} \cite{Necheva97-ARI}. The weak $\gamma$ transition at 167.4~keV of $^{201}$Tl is often buried by the Compton background of the strong 367.9-keV $\gamma$ line from the ($\alpha$,n) reaction product. Owing to this difficulty the lowest measured cross-section point prior to our work was at $E_\alpha = 17.9$~MeV. With the method described in \Sec{sec:analysis} we were able to measure two orders of magnitude lower cross sections down~to~14~MeV.

\subsection{$^{197}$Au($\alpha$,n)$^{200}$Tl}
There are many datasets in the literature for this reaction \cite{Kurz71-NPA, Calboreanu82-NPA, Capurro85-JRNC, Bhardwaj86-NIMA, Shah95-Pra, Necheva97-ARI, Ismail98-Pra, Kulko07-PAN, Basunia07-PRC}. Almost all of these works used the stacked foil technique to measure the reaction cross section at different energies, the only exception is Calboreanu \textit{et al.} \cite{Calboreanu82-NPA}. The energy uncertainty for the reported values are much higher than in this work, where thin single targets were used. All the literature data are in agreement within their error bars; however, either the energy or the cross-section uncertainty is large in the energy range where our investigation has been done. Our new dataset has much higher precision, therefore it provides a better constraint on the theoretical models in this energy region.

\subsection{$^{197}$Au($\alpha$,2n)$^{199}$Tl}
The threshold of the reaction channel with two neutron emission is at $E_\alpha = 17.1$~MeV. Above this energy the cross section increases rapidly. Most of the previously mentioned studies of the ($\alpha$,n) reaction investigated also this reaction channel, and there are a few others \cite{Lanzafame70-NPA, Kurz71-NPA, Calboreanu82-NPA, Capurro85-JRNC, Bhardwaj86-NIMA, Shah95-Pra, Necheva97-ARI, Ismail98-Pra, Kulko07-PAN, Basunia07-PRC}. Similarly to the ($\alpha$,n) channel the literature data are in good agreement within their error bars, but they have limited precision. Our new results are much more precise in the whole investigated energy range, and reach closer to the reaction threshold than ever before.

\section{\label{sec:exp}Experimental details}

\subsection{Targets} Two types of gold targets were used. Either gold layers with typical thicknesses between 0.1 and 0.3~$\mu$m evaporated onto thin aluminum foils or self-supporting gold foils with typical thicknesses of $0.6-0.7$~$\mu$m  were used. The absolute number of target atoms were measured for each target by at least two of the following four independent methods. 

\begin{center}
\begin{figure}[h]
\includegraphics[width=0.32\linewidth]{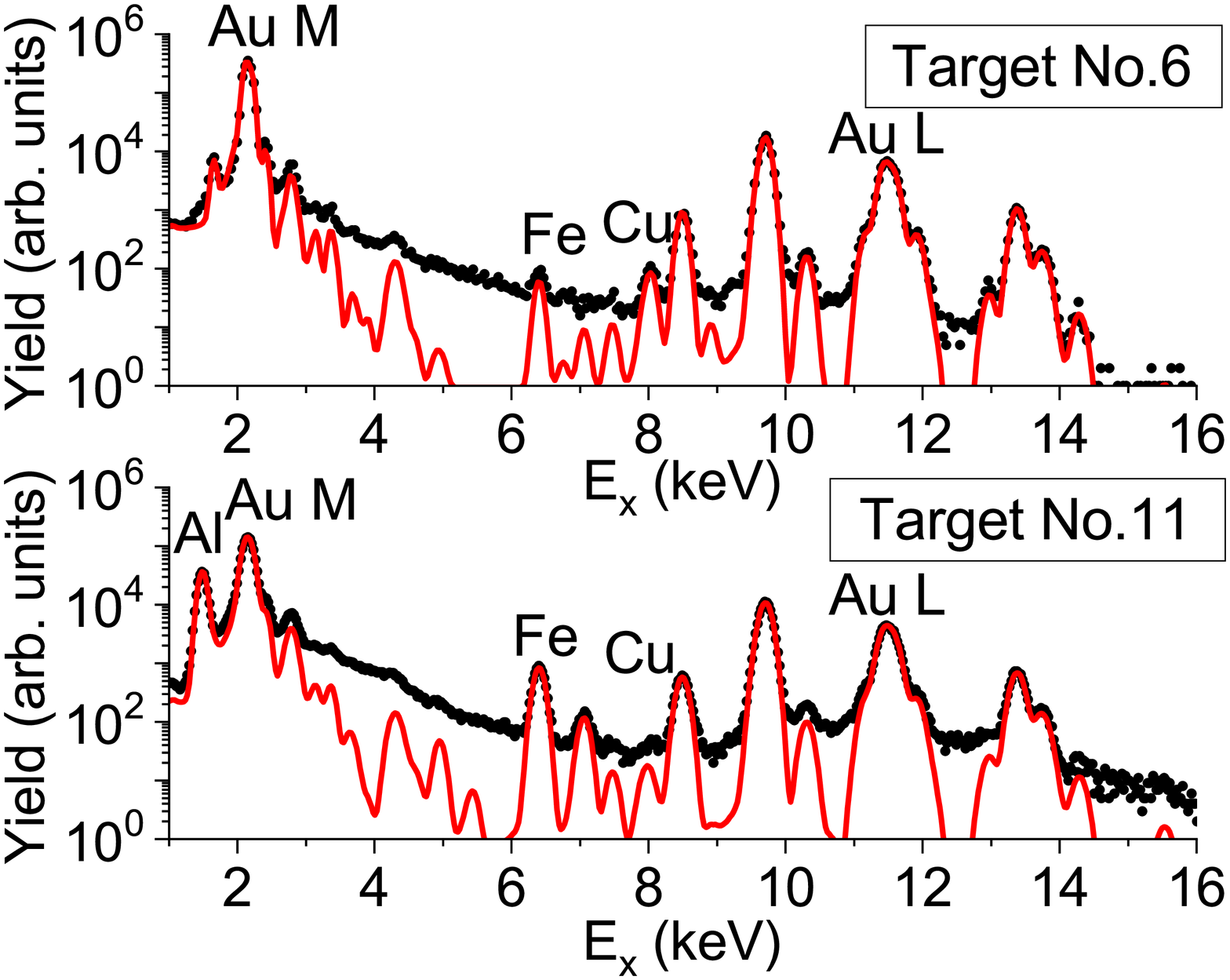}
\includegraphics[width=0.32\linewidth]{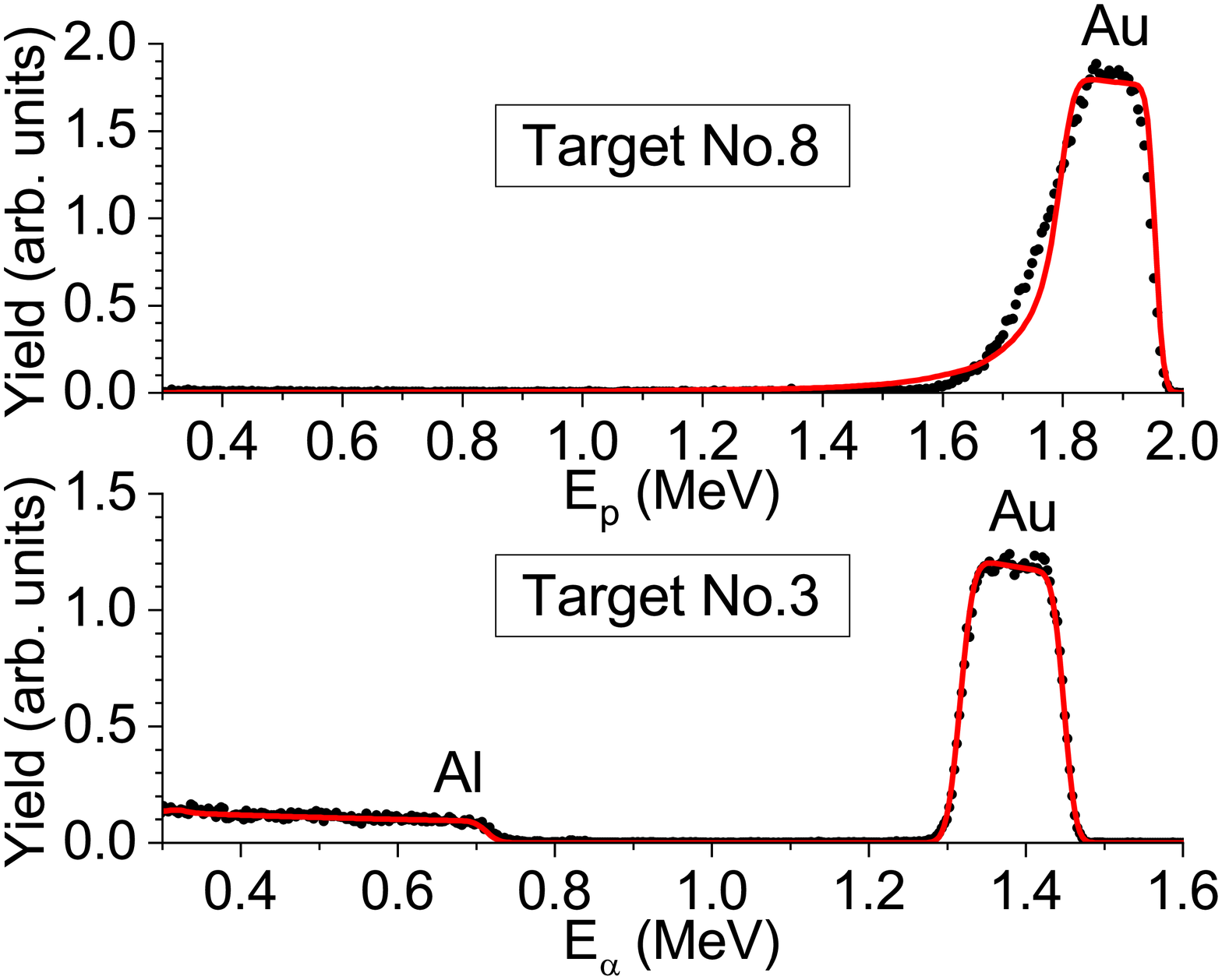}
\includegraphics[width=0.32\linewidth]{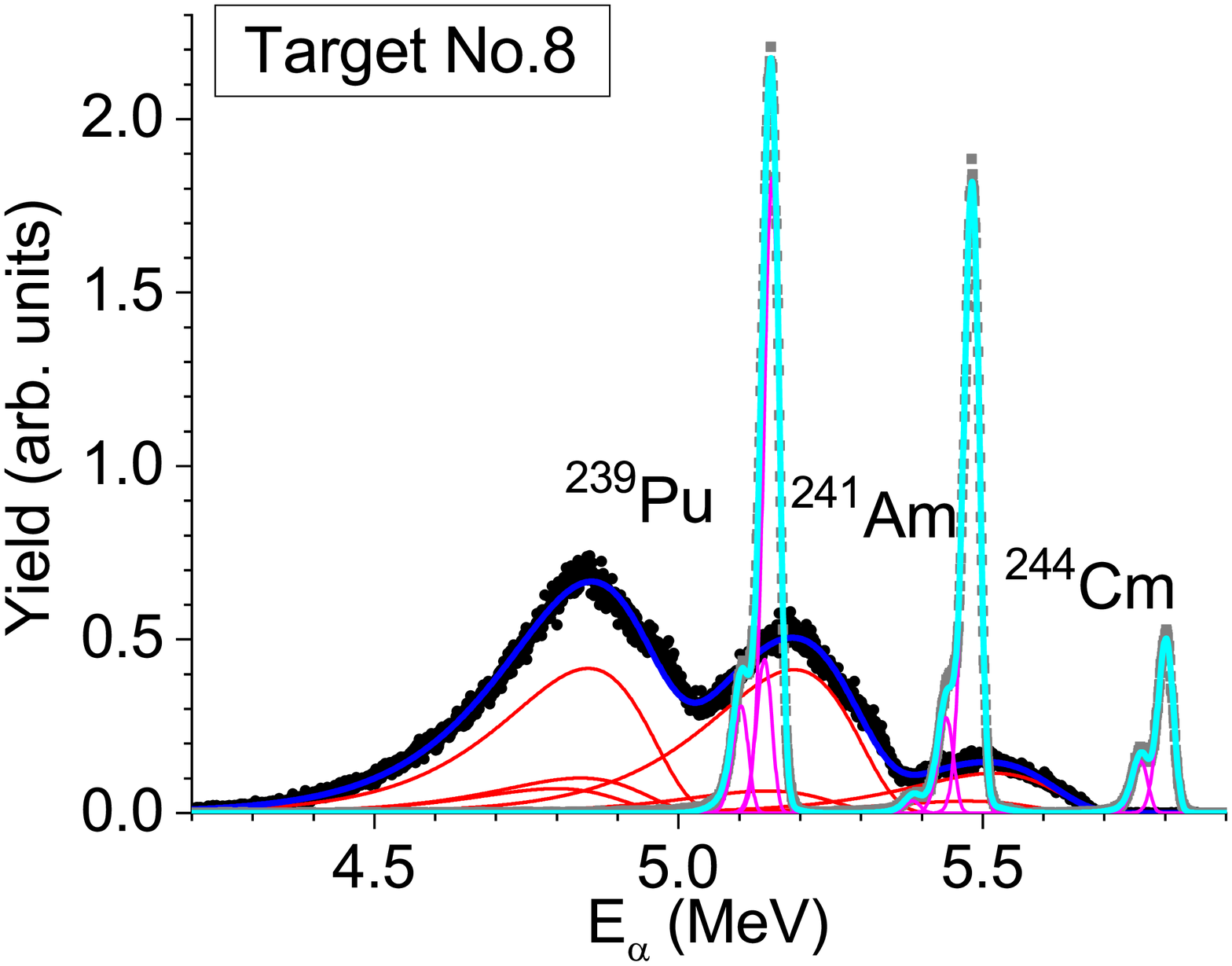}
\caption{\label{fig:thickness} In left column are typical PIXE-spectra in the case of a self-supporting (top) and an evaporated (bottom) target. Middle top figure shows a proton-RBS spectrum of a self-supporting target, while bottom middle is an $\alpha$-RBS spectrum of an evaporated target. Right panel: $\alpha$-spectra of a triple-nuclide $\alpha$-source used for the target thickness measurement is presented, where the black and gray points are the measured energy spectra with and without the gold foil, respectively. The energy calibration fit is plotted with light blue, while the energy loss fit with dark blue lines (see text for details).}
\end{figure}
\end{center}

Both types of targets were investigated by proton induced X-ray emission (PIXE) technique \cite{Koltay11-HNC}. For this purpose the PIXE setup of MTA Atomki installed on the left 45$^\circ$ beamline of the 5-MV Van de Graaff accelerator \cite{Kertesz10-NIMB} was used. A 2-MeV proton beam with about 1- to 3-nA intensity impinged on the targets. Typical PIXE spectra of the two target types can be seen in the left column of \fig{fig:thickness}. The collected spectra were fitted using the GUPIXWIN program code \cite{Campbell10-NIMB}. The final thickness uncertainty of this method is about 4\% including the fit uncertainty and systematic uncertainties concerning the geometry of the setup and the accuracy of the charge measurement.
Besides the thickness determination, the PIXE method allows trace impurity identification in the targets. The self-supporting foils contain Ni and Cu on the 200- and 350-ppm levels, respectively. The aluminum backing of the evaporated targets contains Ti, V, Ni, Cu, Zn, Ga below 50 ppm and Fe with about 3000 ppm.

The targets were investigated also by Rutherford backscattering spectroscopy (RBS) using the Oxford-type Nuclear Microprobe Facility at MTA Atomki \cite{Huszank15-JRNC}.
In the case of the evaporated targets an $\alpha$ beam of 1.6~MeV while for the self-supporting targets a proton beam of 2.0~MeV was used provided by the 5-MV Van de Graaff accelerator. Typical $\alpha$-RBS and proton-RBS spectra are shown in the middle column of \fig{fig:thickness}. The measured RBS spectra were analyzed with the SIMNRA software \cite{SIMNRA}. The uncertainty of the number of target atoms is 3\% from $\alpha$-RBS and 8\%  from proton-RBS. The former is mainly the general accuracy of the given RBS system determined from the measured thickness reproducibility of many standards, and partly from the statistical uncertainty of the fit. The higher uncertainty for the proton-RBS is due to the roughness of the samples causing worse fit.

As a third thickness determination method in the case of the evaporated targets, weighing was used. The weight of each Al foil was measured before and after the evaporation. The target thicknesses were then calculated from the known surface area of the target and the weight difference. The uncertainty of this method is between 2-4\% depending on the thickness of the samples taking into account the precision of the weight measurement (better than 5~$\mu$g) and the possible evaporation non-uniformity.
In the case of the self-supporting foils the energy loss of $\alpha$ particles from a triple-nuclide $\alpha$-source was measured in an ORTEC SOLOIST alpha spectrometer. A typical $\alpha$-energy spectrum is shown in right panel of \fig{fig:thickness}, where the difference between the peak position in the calibration and measurement runs gives the energy loss. The total fit plotted by dark blue and light blue is the sum of the fits made for each of the eight $\alpha$ energies from the source (purple and red lines in the figure). Using the known stopping power, the foil thickness was determined with an accuracy of about 8\% stemming mainly from the stopping power uncertainty (i.\,e. 7.4\%) and partly from that of the measured energy loss.

For each foil the different methods gave consistent results. In the analysis the weighted average of the thickness values obtained with the various methods were used (see \tab{tab:target_and_irrad}).

\onecolumngrid
\vspace{1cm}
\twocolumngrid

\subsection{Irradiations} 
\begin{figure}[t]
\includegraphics[width=0.99\columnwidth]{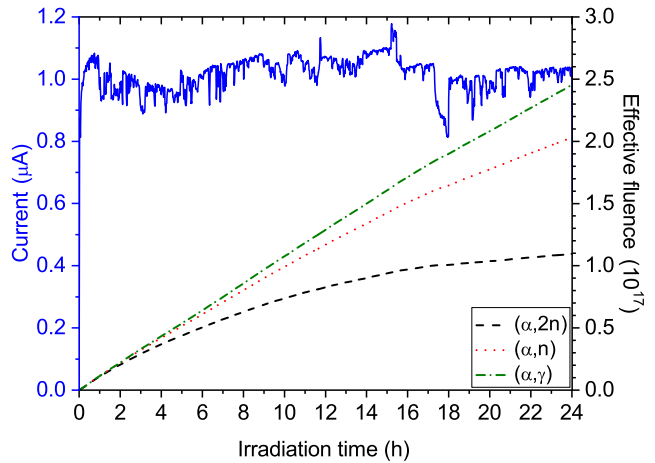}
\caption{\label{fig:irradiation} A typical beam current variation is shown by blue solid line. The time evolution of the effective fluence for the different isotopes are presented by black dashed, red dotted and green dot-dashed lines for $^{199}$Tl, $^{200}$Tl and $^{201}$Tl, respectively.}
\end{figure}

For the irradiations, the MGC-20 type cyclotron of Atomki was used. The $\alpha$ particles entered the activation chamber through a beam defining aperture followed by a second aperture supplied with $-300$~V against secondary electrons either escaping from the chamber or emerging from the collimator. The apertures and the chamber was electrically isolated allowing to measure the beam current. The typical $\alpha^{++}$-beam current was 1 -- 2.5~$\mu$A. The length of the irradiations was typically 20 -- 34~h. 
The beam current was recorded with a multichannel scaler, thus the small variations in the beam intensity could be taken into account in the data analysis.
For the activation analysis described below, we define the effective projectile fluence for each reaction product by the following equation:
\begin{table}[t]
\caption{Target and irradiation parameters. Here the average target thicknesses and the effective beam fluences for each studied isotopes are presented.}
\label{tab:target_and_irrad}
\center
\begin{ruledtabular}
\begin{tabular}{c c c c c }									
\multirow{2}{*}{$E_{\alpha}$ (MeV)}	& \multirow{2}{*}{Target thickness ($\frac{10^{17}\mathrm{at}}{\mathrm{cm}^2}$)}	&	 \multicolumn{3}{c}{$F_{eff}$ ($10^{17}$)} \\
\noalign{\smallskip}\cline{3-5}\noalign{\smallskip}
									&																					&	$^{199}$Tl		& 	$^{200}$Tl	 &	$^{201}$Tl \\
\noalign{\smallskip}\colrule\noalign{\smallskip}								
20.0								& \begin{tabular}{r@{\,$\pm$\,}l} 	42.2	&	1.5	\\	\end{tabular}				&	1.97	&	3.46	&	4.12	\\
19.5								& \begin{tabular}{r@{\,$\pm$\,}l} 	18.66	&	0.24\\	\end{tabular}				&	1.09	&	2.02	&	2.44	\\
19.0								& \begin{tabular}{r@{\,$\pm$\,}l} 	5.28	&	0.11\\	\end{tabular}				&	1.12	&	2.28	&	2.88	\\
18.5								& \begin{tabular}{r@{\,$\pm$\,}l} 	5.22	&	0.11\\	\end{tabular}				&	1.00	&	2.03	&	2.54	\\
18.0								& \begin{tabular}{r@{\,$\pm$\,}l} 	7.10	&	0.13\\	\end{tabular}				&	1.45	&	2.58	&	3.06	\\
17.5								& \begin{tabular}{r@{\,$\pm$\,}l} 	10.27	&	0.16\\	\end{tabular}				&	1.53	&	2.75	&	3.29	\\
17.0								& \begin{tabular}{r@{\,$\pm$\,}l} 	17.86	&	0.23\\	\end{tabular}				&			&	1.88	&	2.34	\\
16.0								& \begin{tabular}{r@{\,$\pm$\,}l} 	43.2	&	2.6	\\	\end{tabular}				&			&	1.88	&	2.26	\\
15.0								& \begin{tabular}{r@{\,$\pm$\,}l} 	37.1	&	1.2	\\	\end{tabular}				&			&	3.31	&	4.15	\\
14.0								& \begin{tabular}{r@{\,$\pm$\,}l} 	41.2	&	1.4	\\	\end{tabular}				&			&	3.29	&	3.95	\\
13.7\footnote{Behind energy degrading foil.}					& \begin{tabular}{r@{\,$\pm$\,}l} 	7.72	&	0.14\\	\end{tabular}				&  			&	3.29	&			\\							
\end{tabular}									
\end{ruledtabular}						
\end{table}
%
\begin{equation}
\label{eq:flux}
F_{eff_x} = \sum_{i=1}^{n}\left(\phi_i\,e^{-(n-i)\,\lambda_x\,\Delta t} \right), 
\end{equation}
where the sum is over each step of the multichannel scaler assuming constant flux ($\phi_i$) within a single time interval of length $\Delta t$ (1~min in this case). $\lambda_x$ is the decay constant of the given isotope (i.\,e. $^{199}$Tl; $^{200}$Tl; $^{201}$Tl). Typical beam current and effective fluence curves as a function of time are shown in \fig{fig:irradiation} and the final effective fluence for each irradiation is presented in \tab{tab:target_and_irrad}.

\newpage
\subsection{$\gamma$-ray and X-ray detection}
The produced activity was determined by counting the $\gamma$ and/or X rays following the decay of the reaction products (see \tab{tab:param}).
In the case of the X-ray counting only the 70.81-keV K$_{\alpha_1}$ X-ray line was used, because the other strong K$_{\alpha_2}$  line at 68.894~keV has a contribution from the X-ray fluorescence peak of gold at 68.806~keV. These two peaks were not separable in the spectrum.

For the counting a thin crystal high-purity germanium detector, a so-called Low Energy Photon Spectrometer (LEPS) was used. The detector was equipped with a home made quasi 4$\pi$ shielding consisting of layers of copper, cadmium and lead \cite{Szucs14-AIPConf}.

\begin{figure}[t]
\includegraphics[width=0.99\columnwidth]{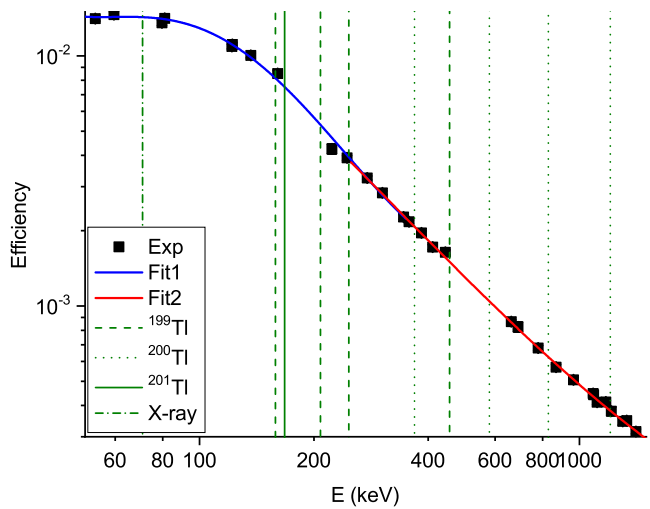}
\caption{\label{fig:eff} The measured detector efficiency at 10~cm (black point) and the fit functions (blue and red curves). Vertical lines indicate the energy of the $\gamma$ rays and the X ray used for the analysis.}
\end{figure}

The detector efficiency calibration was done with $\gamma$ sources of known activity at a counting distance of 10~cm, thus minimizing the true-coincidence summing effect. Since the energies of the decay radiation are between the energies of $\gamma$ rays of the calibration sources, only interpolation was necessary. This was done by fitting log-log polynomial functions to the measured efficiency points. Between 50~keV and 350~keV a 5th order polynomial, while between 250~keV and 1400~keV a 3th order function describes well the measured efficiency. In the overlapping region the two functions are in fair agreement as shown in \fig{fig:eff}.
For the relative efficiency uncertainty, the 1$\sigma$ confidence band of the fits was used.

\begin{figure*}[t]
\includegraphics[width=0.99\linewidth]{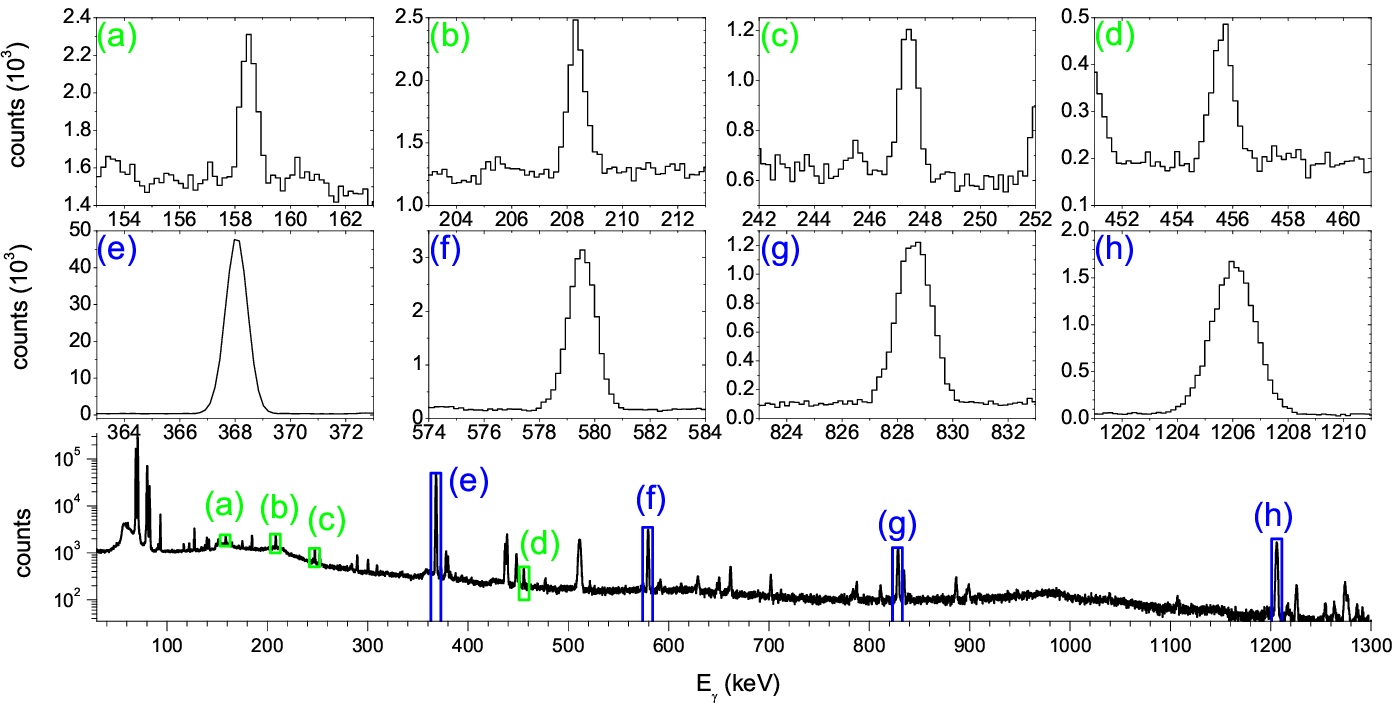}
\caption{\label{fig:an_spe} 1-h long spectrum taken 10~h after the 18.0~MeV irradiation. Insets (a)-(d) show the zoomed regions around the peaks used for the  for $^{199}$Tl activity determination, while (e)-(h) those for $^{200}$Tl.}
\end{figure*}
The targets with lower activity were counted in 3-cm target to detector end-cap distance. At this distance the $\gamma$-ray detection efficiencies were determined by using several targets which were counted both in 10-cm and 3-cm geometry and from the observed count rates, knowing the half-lives of the products and the time difference of the countings, the efficiency-conversion factors were derived. This factor contains the possible loss due to the true-coincidence summing in close geometry. The conversion factors measured with the different sources were consistent, therefore their statistically weighted average was used in the analysis. 

The close-geometry efficiency uncertainty contains the uncertainty of the fit and the uncertainty of the conversion factors, thus ranges from $3-8$\%. The highest values are for the lines of $^{199}$Tl, because they sit on the Compton-continuums of the slower decaying lines of $^{200}$Tl, causing higher statistical uncertainty.
The X-ray detection efficiency in close geometry was determined using the target irradiated with 20.0~MeV.  It was measured two times both at the 10-cm and 3-cm geometry. The counting times were optimized so that for the first counting pair the  $^{200}$Tl, for the second counting pair the $^{201}$Tl dominated the X-ray peak. This was necessary because of the different summing effects characterizing the two isotopes, which lead to slightly different close geometry efficiency of the X ray with the same energy.

Self-absorption effects could be neglected because of the relatively thin targets in the present work. The energy of the detected X ray is just below the K absorption-edge of gold, thus experiences a few times higher absorption as the $\gamma$ rays. For the thickest gold target (see Table II) the X-ray self-absorption is less than 0.2\% considering an even activity distribution in the target, thus can be neglected safely.

\section{\label{sec:analy}Data analysis}
\subsection{($\alpha$,2n) and ($\alpha$,n) cross sections}

\begin{figure*}[t]
\includegraphics[width=0.49\linewidth]{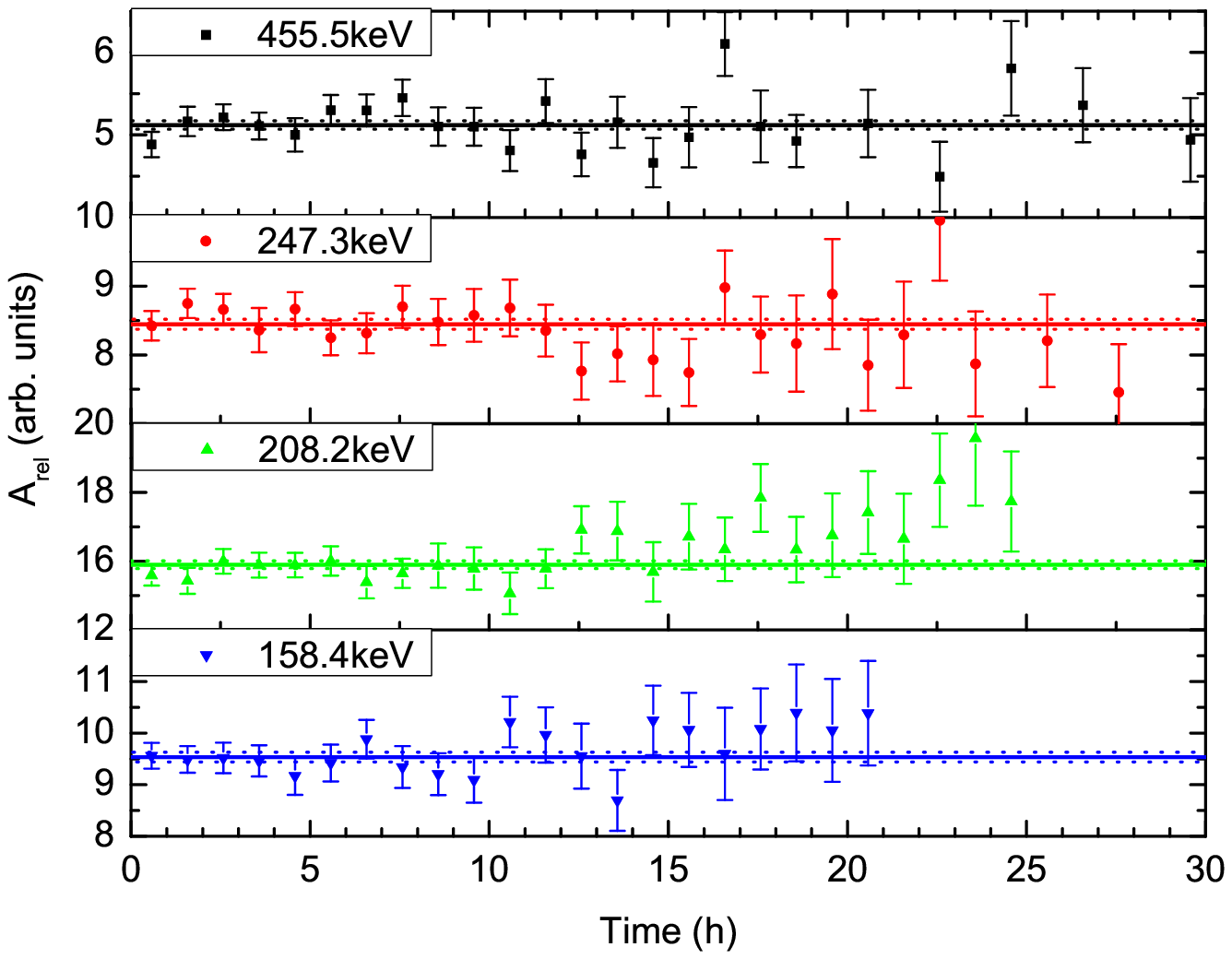}
\includegraphics[width=0.49\linewidth]{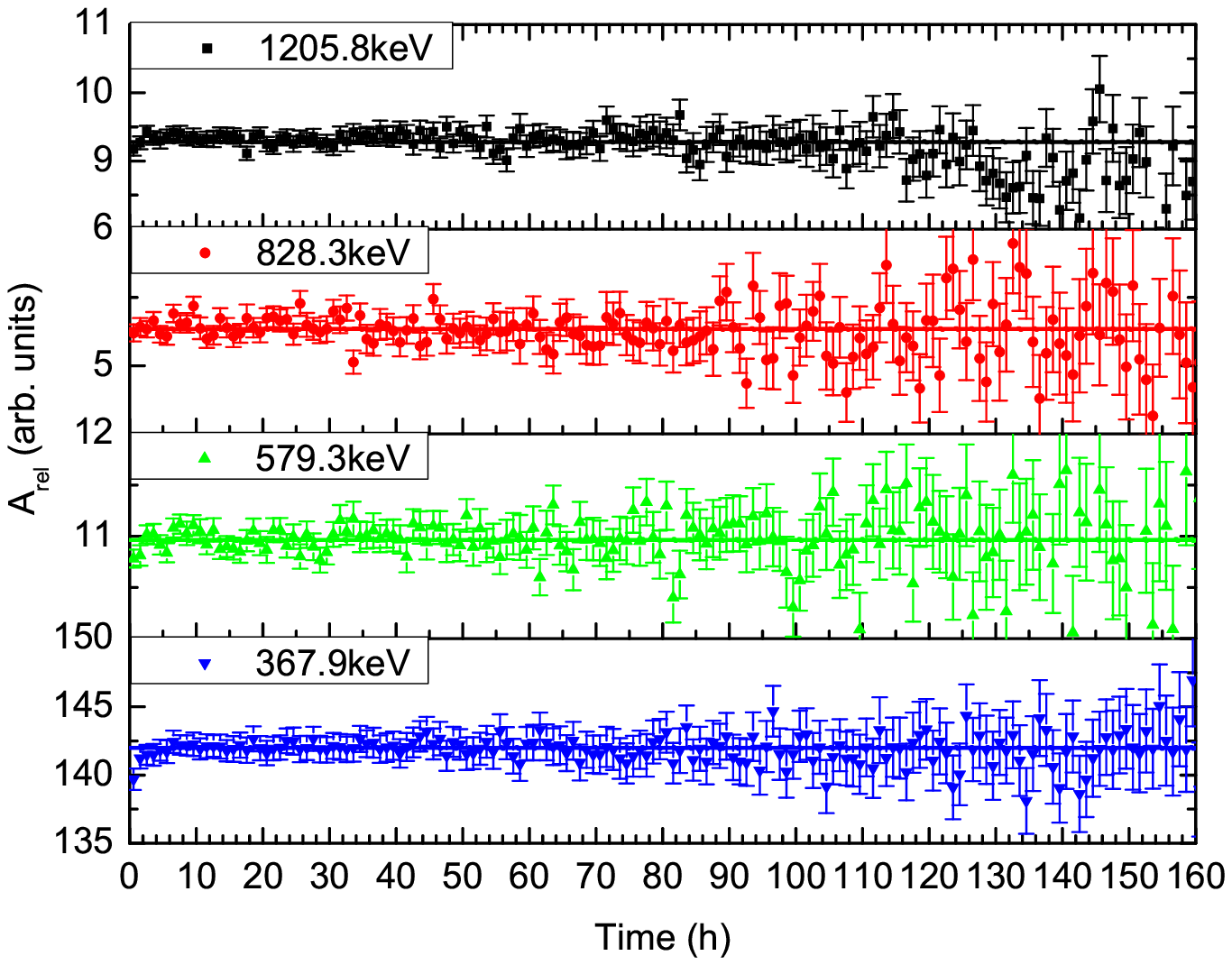}
\caption{\label{fig:rel_act_an} Relative activity as a function of time. The dots are the measured values for a given transition. Horizontal solid lines are the average values, while dotted lines indicate the uncertainty of the average (if it is smaller than the line width of the average, then the lines are not shown). The left and right panels are for the transitions of the ($\alpha$,2n) and ($\alpha$,n) reaction products, respectively.}
\end{figure*}

First the activity of the ($\alpha$,n) and above 17.1~MeV also that of the ($\alpha$,2n) reaction products were determined as follows.
The $\gamma$ peaks were fitted with Gaussian plus linear background in each hourly-recorded spectra (see \fig{fig:an_spe}). If the uncertainty of the peak area from the fit was more than 10\% then spectra were added together until the fit resulted in lower than 10\% statistical uncertainty. This spectrum summing was done separately for each of the studied peaks.

The peak areas were then divided by the corresponding waiting and counting time factors, yielding a number related to the activity of the given isotope at the end of the irradiation (hereafter referred to as the relative activity). The statistically weighted average of these individual numbers (see e.\,g. \fig{fig:rel_act_an}) is the finally obtained relative activity ($A_{rel}$), which was calculated with the following equation: 
\begin{equation}
\label{eq:rel_act}
A_{rel_x} = \left( \sum_{i=1}^{n} \frac{C_i}{e^{-\lambda_x t_{w_i}} \left( 1 - e^{-\lambda_x t_{c_i}}  \right)} W_i \right) \bigg/ \left( \sum_{i=1}^{n}  W_i \right),
\end{equation}
where the summation runs till the last counting which results in better than 10\% fit uncertainty. $C_i$ is the peak area from the $i$-$th$ counting, $\lambda_x$ is the decay constant of isotope $x$, $ t_{w_i}$ and $t_{c_i}$ are the waiting time and length of the $i$-$th$ counting, respectively. The weighting factors $W_i$ are the square of the reciprocal statistical uncertainty coming mainly from the fitted peak areas and partly from the uncertainty of the counting and waiting factors. The uncertainty of the relative activities are the reciprocal square-root of the sum of the weights.
The relative activity determination for the irradiation at 18.0~MeV is plotted in \fig{fig:rel_act_an}.

The relative activities of a given transition were then divided by the relative intensities (${I_{rel_i}}$) and detection efficiencies ($\eta_i$) of the corresponding $\gamma$ rays. Consistent values were obtained for all the studied transitions, thus the created activity of a given isotope at the end of the irradiation was calculated as follows: 
\begin{equation}
A_x = \left(\left( \sum_{i=1}^{n} \frac{A_{rel_i}}{I_{rel_i} \eta_i} w_i \right) \bigg/ \left(\sum_{i=1}^{n}  w_i \right) \right) \Bigg/ M,
\end{equation}
where the summation goes over the four transition of the isotope in question.
The weights are formed from the combined uncertainty of $A_{rel_i}$, ${I_{rel_i}}$ and $\eta_i$. Finally the average is divided by the multiplicator for the absolute intensity $M$.

At the very end the reaction cross sections are obtained by the following equation:
\begin{equation}
\sigma = \frac{A_x}{D F_{eff_x}},
\end{equation}
where $A_x$ is the created activity of isotope $x$ in the sample, $D$ is the target thickness and $F_{eff_x}$ is the effective irradiation fluence as defined in \eq{eq:flux}.

\subsection{\label{sec:analysis} ($\alpha$,$\gamma$) cross sections}

\begin{figure}[t]
\includegraphics[width=0.99\columnwidth]{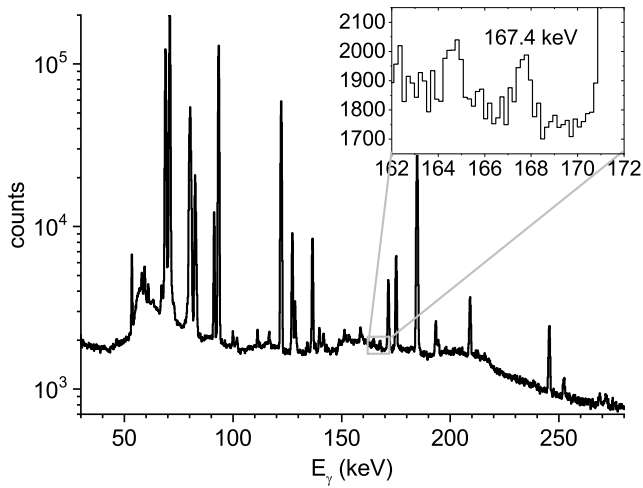}
\caption{\label{fig:ag_spe_g} 101-h long spectrum collected 150~h after the 17.5-MeV irradiation. The inset shows the $\gamma$ peak of $^{201}$Tl.}
\end{figure}

The $\gamma$ rays from the decay of $^{201}$Tl were only visible at and above 17.5-MeV bombarding energy (see \fig{fig:ag_spe_g}), because of the Compton background of the very intense 367.9-keV $\gamma$ line of the ($\alpha$,n) reaction product and other parasitic activities created on the trace impurities of the targets. Owing to the common systematic uncertainties of the $\gamma$-ray and X-ray counting methods, the final uncertainty would not decrease by averaging. Therefore the adopted cross section was only derived from the X-ray counting as it has much higher precision.
The produced $^{201}$Tl activity from the X-ray counting was determined as follows.

\begin{figure}[b]
\includegraphics[width=0.99\columnwidth]{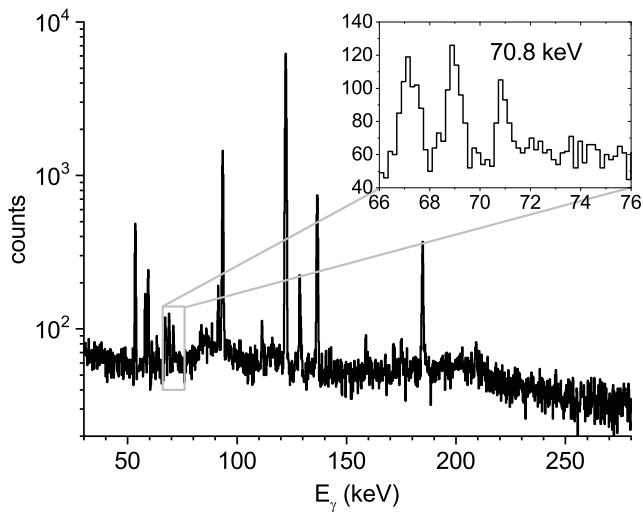}
\caption{\label{fig:ag_spe_X} 11-h long spectrum collected 453~h after the 17.5-MeV irradiation. The inset shows the X-ray peak of $^{201}$Tl, together with the X-ray fluorescence peaks of gold. Comparing this spectrum to the one in \fig{fig:ag_spe_g} the higher sensitivity of the X-ray counting method is clearly seen.}
\end{figure}

The X-ray peak was fitted by Gaussian and quadratic background. Similar to the $\gamma$-peak fits, spectra were added together so that the peak fit in this case resulted in less than 20\% peak area uncertainty.

Owing to the half-life difference, after about 16 days, the ($\alpha$,$\gamma$) reaction product $^{201}$Tl had a large enough contribution to the X-ray peak. The subtraction of the contribution from the other two reaction products became possible, as discussed below. A typical spectrum used for the X-ray activity determination is shown in \fig{fig:ag_spe_X} where already more than half of the peak counts are caused by the $^{201}$Tl activity.

\begin{figure}[b]
\includegraphics[width=0.99\columnwidth]{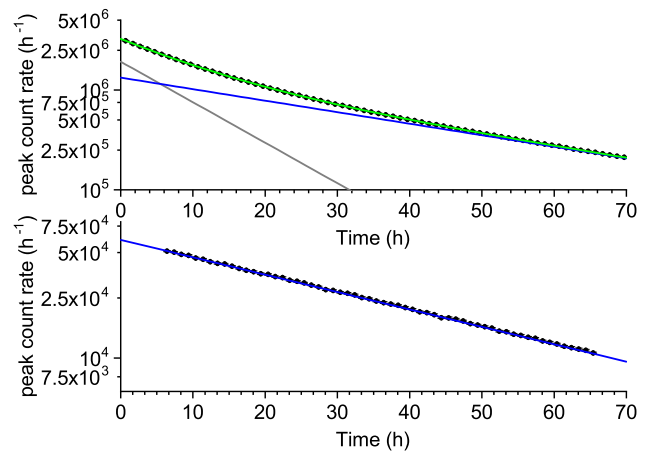}
\caption{\label{fig:decay} Time evolution of the count rate in the X-ray peak (black dots). The upper panel shows the dual exponential fit after the 19.5-MeV irradiation. Gray, blue and green solid lines represent the fitted $^{199}$Tl and $^{200}$Tl contribution and their sum, respectively. Lower panel shows a single exponential fit in the case of the 17.0-MeV measurement, where the blue line is the fitted exponential with the $^{200}$Tl half-life.}
\end{figure}

The subtraction of the contributions from the other isotopes was done as follows.
Below the ($\alpha$,2n) reaction threshold the X-ray decay-curve recorded in the first days after the irradiation was fitted using the $^{200}$Tl half-life, thus the X-ray relative activity of $^{200}$Tl was determined (see lower panel of \fig{fig:decay}). 
Above the ($\alpha$,2n) reaction threshold, the X-ray decay curve was fitted with the sum of two exponentials with the known half-lives of $^{199}$Tl and $^{200}$Tl, similarly as e.g. in Kiss \textit{et al.} \cite{Kiss18-PRC}. From the fit the relative X-ray activity for both reaction products was derived. An example of such a fit is also shown in \fig{fig:decay} (upper panel).
In the first days of the countings the contribution of the ($\alpha$,$\gamma$) reaction product is negligible to the X-ray peak as calculated using the literature X-ray intensities and the produced activity previously determined via $\gamma$ counting.
\begin{figure}[b]
\includegraphics[width=0.99\columnwidth]{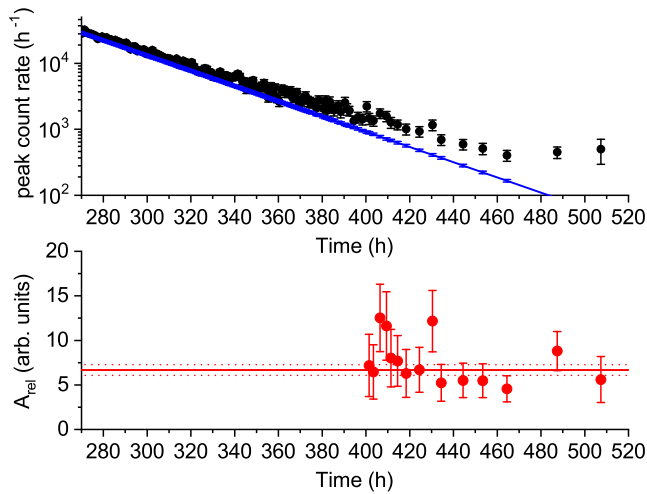}
\caption{\label{fig:subtr} Upper panel shows the time evolution of the count rate in the X-ray peak (black dots) together with the calculated contribution from the ($\alpha$,n) reaction channel (blue line with error bars) in the case of the last 150 hours of counting of the 17.5-MeV sample. The lower panel shows the relative activity from the X-ray measurement of $^{201}$Tl. Red dots are the decay and counting corrected peak areas after subtraction. The subtraction was possible with reasonable accuracy only after 400 hours.}
\end{figure}

The X-ray countings for the $^{201}$Tl activity were done at least 15~days after the irradiations. After such a waiting time the contribution from the ($\alpha$,2n) reaction product was always negligible.
For the subtraction of the X-ray contribution of the ($\alpha$,n) reaction product, the relative activity ratio of the 367.9-keV $\gamma$ ray and the X ray was used. The ratio was determined from several samples at the actual 3-cm counting geometry. Since the uncertainty of the relative activity from the 367.9-keV $\gamma$ line contains only the uncertainty of the counting statistics, the number to be subtracted is more precise than what would be calculated from the absolute activity. This latter would contain the uncertainties of the detection efficiencies and X-ray branchings.

After subtracting the ($\alpha$,n) contribution, only those points were used where the relative uncertainty of the remaining peak areas was not higher than 50\%. Then those were corrected for the decay and counting time resulting in the X-ray relative-activity values. The final relative activity is the weighted average of those from the subsequent countings, similar to the $\gamma$-ray relative activities (see \fig{fig:subtr}).

The 13.7-MeV point was measured differently from the others. For the 14-MeV irradiation two targets were placed in the irradiation chamber separated by a 2.13~$\mu$m thick Al foil. The beam energy at the position of the second target is calculated using SRIM, from the known thickness of the first target and the Al energy degrader foil. The energy uncertainty of this point is therefore higher.
At this energy the 367.9-keV $\gamma$ line from the ($\alpha$,n) reaction product was not visible during the course of the 1.5-days counting right after the irradiation. Therefore,
 the activity was determined from X-ray counting only using the absolute X-ray branching ratio from the literature (see \tab{tab:param}). The X-ray peak count rate followed the half life of the $^{200}$Tl, thus was considered to be populated only by the ($\alpha$,n) reaction product.

In the case of the self-supporting foils having no backing, some activity can be lost when the reaction takes place at the rear of the target layer, and the reaction product cannot be stopped in the remaining part of the foil. A SRIM simulation \cite{srim} was done to estimate this effect. As a starting point of the simulation, Tl nuclei were equally distributed in the gold foil, and each of them had a velocity directing to the rear of the foil, calculated from the reaction kinematics for each irradiation energy.
The simulation showed that about 3\% of the Tl nuclei can leave the gold foils. This loss was finally taken into account in the created activity determination with a conservative 30\% relative uncertainty.

The effective energy in each case is taken at the middle of the target. The energy loss is calculated with SRIM \cite{srim}.
The energy uncertainty determined by the initial beam-energy uncertainty of 0.3\%. The effect of the target thickness and energy loss on beam-energy uncertainty is $0.005-0.05$\%, thus neglected.

Beside the statistical uncertainties propagated with the averaging as discussed before, each data point contains the respective target thickness uncertainty as quoted in \tab{tab:target_and_irrad} and the uncertainty of the Tl loss in the case of the self-supporting targets. As systematic uncertainty, the absolute branching (7-10\%), the absolute detection efficiency (3\%) and the beam current (3\%) uncertainty was quadratically added to get the finally quoted uncertainties.
The absolute detection efficiency uncertainty accounts for the uncertainty of the absolute activity of the calibration sources and that of the counting distance reproducibility.

\section{\label{sec:res} Experimental results}
\subsection{\label{sec:branching} X-ray intensities}

\begin{table}[b]
\caption{Relative intensity of the K$_{\alpha_1}$ X rays to the strongest $\gamma$ ray (marked with 100 in \tab{tab:param}) for several runs. In the case of $^{201}$Tl the statistic of the $\gamma$ ray was sufficient for the analysis only at 20~MeV. The uncertainty of the averaged value includes the relative detection efficiency uncertainty.}
\label{tab:X_branch}
\center
\begin{ruledtabular}
\begin{tabular}{l c c c }
E$_{\alpha}$ (MeV)	&	 \multicolumn{3}{c}{Relative X-ray intensity (\%)}	\\
\noalign{\smallskip}\cline{2-4}\noalign{\smallskip}
					& 									$^{199}$Tl					& 									$^{200}$Tl						&										 $^{201}$Tl \\ 
\noalign{\smallskip}\colrule\noalign{\smallskip}	
20.0				& \begin{tabular}{r@{\,$\pm$\,}l}	322	&	15	\\ \end{tabular} & \begin{tabular}{r@{\,$\pm$\,}l}	43.5	&	0.2	\\ \end{tabular} & \begin{tabular}{r@{\,$\pm$\,}l}	405		&	40	\\ \end{tabular} \\
19.5				& \begin{tabular}{r@{\,$\pm$\,}l}	346	&	7	\\ \end{tabular} & \begin{tabular}{r@{\,$\pm$\,}l}	43.2	&	0.4	\\ \end{tabular}	& 																	\\
19.0				& \begin{tabular}{r@{\,$\pm$\,}l}	347	&	8	\\ \end{tabular}	& \begin{tabular}{r@{\,$\pm$\,}l}	43.8	&	0.3	\\ \end{tabular} & 																	\\
17.0				& 																& \begin{tabular}{r@{\,$\pm$\,}l}	43.2	&	0.8	\\ \end{tabular} & 																	\\
\noalign{\smallskip}\colrule\noalign{\smallskip}	
Average				& \begin{tabular}{r@{\,$\pm$\,}l}	344	&	6	\\ \end{tabular} & \begin{tabular}{r@{\,$\pm$\,}l}	43.5	&	0.5	\\ \end{tabular} & \begin{tabular}{r@{\,$\pm$\,}l}	405		&	41	\\ \end{tabular} \\
\noalign{\smallskip}\colrule\noalign{\smallskip}	
Ref.\cite{Debertin79-ARI} &	&	& \begin{tabular}{r@{\,$\pm$\,}l}	446		&	12	\\ \end{tabular}
\end{tabular}
\end{ruledtabular}
\end{table}

The X-ray relative activity of all the created isotopes is determined for several samples. Using the $\gamma$-ray relative activity and the X- and $\gamma$-ray detection efficiencies, the relative X-ray branching ratios are determined here.
For each isotopes \tab{tab:X_branch} presents the X-ray branching ratios relative to the strongest $\gamma$ lines (i.e. 455.5~keV for $^{199}$Tl, 367.9~keV for $^{200}$Tl and 167.4~keV for $^{201}$Tl, respectively). 
To avoid the systematic effect of the efficiency scaling, only data points measured at the 10-cm counting geometry are used. The uncertainty of the relative detection efficiency was added to the final value after averaging.
A comparison with other measured relative branching ratios is possible only in the case of $^{201}$Tl. For the other isotopes no published values are available in the literature.

The absolute intensities were calculated by scaling the relative values by the multiplicator shown in \tab{tab:param}. For each isotope the measured absolute intensities can be compared to the values presented in NuDat2 \cite{NuDat}. In the  NuDat2 database the X-ray branching ratios are obtained with the RADLIST \cite{RADLIST} program using the internal conversion coefficients. The present experimental data found to be in agreement with the calculated values from the database (see \tab{tab:X_abs_branch}). In the case of $^{199}$Tl and $^{200}$Tl the experiential values are somewhat less precise owing to the uncertainty of the multiplicator. However, for $^{201}$Tl the precision limiting factor was the counting statistics, the obtained branching ratio is more precise than that in the database. For this isotope the latest evaluation also contains experimental data for the X-ray intensities. The new value has to be compared with the more precise evaluated value of 44.6\,$\pm$\,0.6\% \cite{NDS201} stated also in \tab{tab:X_branch} and used for the $^{197}$Au($\alpha$,$\gamma$)$^{201}$Tl cross section determination.

\subsection{\label{sec:XS} Reaction cross sections}

The measured reaction cross sections are shown in \tab{tab:XS}. In the case of the 18.5-MeV data point only $\gamma$ counting was done. In this measurement the $\gamma$ peak from the $^{197}$Au($\alpha$,$\gamma$)$^{201}$Tl reaction was not visible, thus no cross section could be derived. The total uncertainties presented in the table are the quadratic sum of the systematic uncertainties (10.6\%, 8.1\%, and 4.5\% for the ($\alpha$,2n), ($\alpha$,n), and ($\alpha$,$\gamma$) reactions respectively) and statistical uncertainties of the datapoints. The latter varies between 2-6\% for the neutron emitting reactions while for the radiative-capture reaction it is 9-15\% except for the two lowest energy data points (26\% and 54\%).
  
\begin{table}[t]
\caption{Absolute X-ray intensities in \% determined in the present work and compared to their values from the NuDat2 \cite{NuDat} database.}
\label{tab:X_abs_branch}
\center
\begin{ruledtabular}
\begin{tabular}{l c c }
Isotope		& 								NuDat2 \cite{NuDat}					& 									This work	\\
\noalign{\smallskip}\colrule\noalign{\smallskip}	
$^{199}$Tl	& \begin{tabular}{r@{\,$\pm$\,}l}	45.5&	2.5	\\ \end{tabular} & \begin{tabular}{r@{\,$\pm$\,}l}	42.7	&	4.2	\\ \end{tabular} \\
$^{200}$Tl	& \begin{tabular}{r@{\,$\pm$\,}l}	40.4&	1.7	\\ \end{tabular} & \begin{tabular}{r@{\,$\pm$\,}l}	37.8	&	2.6	\\ \end{tabular}	\\
$^{201}$Tl	& \begin{tabular}{r@{\,$\pm$\,}l}	37  &	6	\\ \end{tabular}	& \begin{tabular}{r@{\,$\pm$\,}l}	40.5	&	4.1	\\ \end{tabular} 	\\
\end{tabular}
\end{ruledtabular}
\end{table}

\section{Theoretical analysis}
\label{sec:theo}
\subsection{Formalism and general remarks}
\label{sec:form}
The new experimental data were analyzed within the statistical model (SM). In a schematic notation, the cross section $\sigma$\raX\ of an \al -induced reaction is given by
\begin{equation}
\sigma(\alpha,X) \sim \frac{T_{\alpha,0} T_X}{\sum_i T_i} = T_{\alpha,0}
\times b_X
\label{eq:StM}
\end{equation}
with the transmission coefficients $T_i$ into the $i$-th open channel and the branching ratio $b_X = T_X / \sum_i T_i$ for the decay into the channel $X$. The total transmission is given by the sum over all contributing channels: $T_{\rm{tot}} = \sum_i T_i$. The $T_i$ are calculated from global optical potentials for the particle channel and from the $\gamma$-ray strength function (GSF) for the photon channel. The $T_i$ include contributions of all final states $j$ in the respective residual nucleus in the $i$-th exit channel. In practice, the sum over all final states $j$ is approximated by the sum over low-lying excited states up to a certain excitation energy $E_{\rm{LD}}$ (these low-lying levels are typically known from experiment) plus an integration over a theoretical level density for the contribution of higher-lying excited states:
\begin{equation}
T_i = \sum_j T_{i,j} \approx 
\sum_j^{E_j < E_{\rm{LD}}} T_{i,j} +
\int_{E_{\rm{LD}}}^{E_{\rm{max}}} \rho(E) \, T_i(E) \, dE
\label{eq:Tsum}
\end{equation}
\onecolumngrid
\begin{center}
\begin{table}[h]
\caption{Measured reaction cross sections with their total uncertainties.}
\label{tab:XS}
\center
\begin{ruledtabular}
\begin{tabular}{l c c  c c }
E$_{\alpha}$ (MeV)	&	E$_{eff}$ (MeV)											& 				$^{197}$Au($\alpha$,2n)$^{199}$Tl (mb)			& 					$^{197}$Au($\alpha$,n)$^{200}$Tl (mb)						&
	$^{197}$Au($\alpha$,$\gamma$)$^{201}$Tl ($\mu$b) \\ 
\noalign{\smallskip}\colrule\noalign{\smallskip}	
20.0	& \begin{tabular}{r@{\,$\pm$\,}l}	19.92	&	0.06	\\ \end{tabular} & \begin{tabular}{r@{\,$\pm$\,}l}	36		&	4	\\ \end{tabular} & \begin{tabular}{r@{\,$\pm$\,}l}	37.0	&	3.3		\\ \end{tabular} & \begin{tabular}{r@{\,$\pm$\,}l}	20.5	&	2.1	\\ \end{tabular} \\
19.5	& \begin{tabular}{r@{\,$\pm$\,}l}	19.46	&	0.06	\\ \end{tabular} & \begin{tabular}{r@{\,$\pm$\,}l}	16.6	&	1.8	\\ \end{tabular} & \begin{tabular}{r@{\,$\pm$\,}l}	25.2	&	2.1		\\ \end{tabular} & \begin{tabular}{r@{\,$\pm$\,}l}	15.6	&	1.7	\\ \end{tabular} \\
19.0	& \begin{tabular}{r@{\,$\pm$\,}l}	18.99	&	0.06	\\ \end{tabular} & \begin{tabular}{r@{\,$\pm$\,}l}	6.3		&	0.7	\\ \end{tabular} & \begin{tabular}{r@{\,$\pm$\,}l}	17.6	&	1.5		\\ \end{tabular} & \begin{tabular}{r@{\,$\pm$\,}l}	15.4	&	1.8	\\ \end{tabular} \\
18.5	& \begin{tabular}{r@{\,$\pm$\,}l}	18.49	&	0.06	\\ \end{tabular} & \begin{tabular}{r@{\,$\pm$\,}l}	1.80	&	0.20\\ \end{tabular} & \begin{tabular}{r@{\,$\pm$\,}l}	10.6	&	0.9		\\ \end{tabular} & 																				\\
18.0	& \begin{tabular}{r@{\,$\pm$\,}l}	17.99	&	0.05	\\ \end{tabular} & \begin{tabular}{r@{\,$\pm$\,}l}	0.34	&	0.04\\ \end{tabular} & \begin{tabular}{r@{\,$\pm$\,}l}	6.0		&	0.5		\\ \end{tabular} & \begin{tabular}{r@{\,$\pm$\,}l}	5.9		&	0.8	\\ \end{tabular} \\
17.5	& \begin{tabular}{r@{\,$\pm$\,}l}	17.48	&	0.05	\\ \end{tabular} & \begin{tabular}{r@{\,$\pm$\,}l}	0.046	&	0.005\\ \end{tabular} & \begin{tabular}{r@{\,$\pm$\,}l}	3.32	&	0.28	\\ \end{tabular} & \begin{tabular}{r@{\,$\pm$\,}l}	3.5		&	0.5	\\ \end{tabular} \\
17.0	& \begin{tabular}{r@{\,$\pm$\,}l}	16.96	&	0.05	\\ \end{tabular} & 																	& \begin{tabular}{r@{\,$\pm$\,}l}	1.28	&	0.11	\\ \end{tabular} & \begin{tabular}{r@{\,$\pm$\,}l}	1.67	&	0.22	\\ \end{tabular} \\
16.0	& \begin{tabular}{r@{\,$\pm$\,}l}	15.91	&	0.05	\\ \end{tabular} & 																	& \begin{tabular}{r@{\,$\pm$\,}l}	0.226	&	0.023	\\ \end{tabular} & \begin{tabular}{r@{\,$\pm$\,}l}	0.45	&	0.07 \\ \end{tabular} \\
15.0	& \begin{tabular}{r@{\,$\pm$\,}l}	14.92	&	0.05	\\ \end{tabular} & 																	& \begin{tabular}{r@{\,$\pm$\,}l}	0.0249	&	0.0022	\\ \end{tabular} & \begin{tabular}{r@{\,$\pm$\,}l}	0.081	&	0.021 \\ \end{tabular} \\
14.0	& \begin{tabular}{r@{\,$\pm$\,}l}	13.91	&	0.04	\\ \end{tabular} & 																	& \begin{tabular}{r@{\,$\pm$\,}l}	0.00141	&	0.00013	\\ \end{tabular} & \begin{tabular}{r@{\,$\pm$\,}l}	0.037	&	0.020 \\ \end{tabular} \\
13.7\footnote{Measured with energy degrader foil. See text for details.}	&  \begin{tabular}{r@{\,$\pm$\,}l} 	13.62	&	0.05	\\ \end{tabular} &								 						& \begin{tabular}{r@{\,$\pm$\,}l}	0.00067	&	0.00007	\\ \end{tabular} & 																				\\
\end{tabular}
\end{ruledtabular}
\end{table}
\end{center}
\vspace{1cm}
\twocolumngrid

\noindent For further details of the definition of $T_i$, see \cite{Rauscher11-IJMPE}. $T_{\alpha,0}$ refers to the entrance channel where the target nucleus is in its ground state under laboratory conditions. The calculation of stellar reaction rates \Nsv\ may require further modifications of Eq.~(\ref{eq:StM}) which have to take into account thermal excitations of the target nucleus \cite{Rauscher11-IJMPE}.

From Eqs.~(\ref{eq:StM}) and (\ref{eq:Tsum}), the following properties of the reactions under study can be expected. The total reaction cross section \stot\ (summed over all non-elastic channels) depends only on the transmission $T_{\alpha,0}$ in the entrance channel and is thus only sensitive to the chosen \al -nucleus optical model potential (A-OMP). At very low energies all particle channels are closed. Here the only open reaction channel is the \rag\ channel, leading to a cross section of the \rag\ reaction of about \stot ; consequently, at very low energies the \rag\ cross section is essentially only sensitive to the chosen A-OMP.

The \rap\ channel opens at about 6.5 MeV. However, because of the high Coulomb barrier in the exit channel, the transmission $T_p$ remains practically negligible in the energy range under study, and a more detailed discussion of the \rap\ channel will be omitted.

Contrary to the \rap\ channel, there is no Coulomb barrier for the \ran\ channel. Already close above the \ran\ threshold at 9.7 MeV, the transmission $T_n$ exceeds all transmissions $T_{X \ne n}$ into other channels. Now  the \ran\ cross section becomes close to the total reaction cross section \stot , and thus the \ran\ cross section is practically only sensitive to the chosen A-OMP. This finding holds until energies of about 17 MeV where the \rann\ channel opens. At these higher energies the total reaction cross section is essentially distributed among the \ran\ and \rann\ channels; i.e., the sum of the \ran\ and \rann\ cross sections is approximately given by \stot\ and is sensitive to the A-OMP only. But the individual \ran\ and \rann\ cross sections are sensitive to the ratio between $T_{n}$ and $T_{2n}$ which in turn depend on the chosen nucleon-nucleus optical model potential (N-OMP) and on the chosen level densities (LD) for the residual \tlnull\ and \tlix\ nuclei.

At all energies above the \ran\ threshold, the \rag\ cross section depends on the ratio $T_{\alpha,0} T_\gamma / T_{\rm{tot}}$ and is thus sensitive not only to the transmission $T_\gamma$ and the $\gamma$-ray strength function, but also sensitive to all further ingredients like the A-OMP, N-OMP, and LD. The analysis of the \rag\ excitation function alone does not allow to fix any ingredient of the SM calculations because of the complex sensitivity of the \rag\ cross section.

\subsection{Additional data from elastic scattering}
\label{sec:elast}
The total reaction cross section \stot\ can also be derived from the analysis of elastic scattering angular distributions. It has been shown recently that \stot\ extracted from elastic scattering is consistent with the sum over the \raX\ cross sections of all non-elastic channels \cite{Gyurky12-PRC,Ornelas16-PRC}. 
\begin{figure}[t]
\includegraphics[width=0.99\columnwidth]{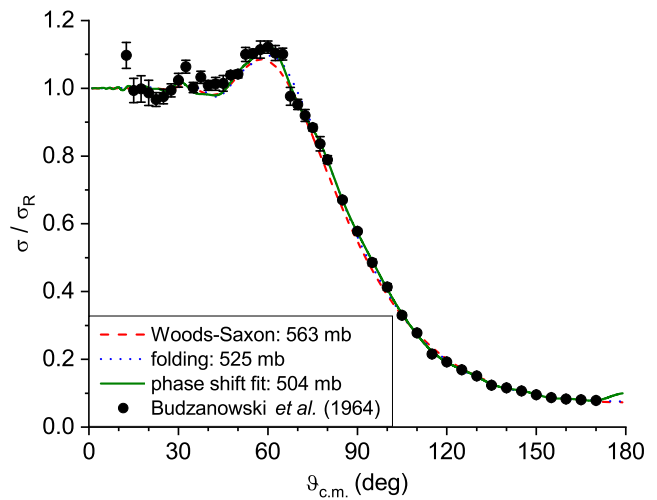}
\caption{
  \label{fig:e25scat}  \auvii \raa \auvii\ elastic scattering at $E_\alpha = 24.7$ MeV: a total   reaction cross section \stot\ = $520 \pm 20$ mb is derived from the angular  distribution of \cite{Budzanowski64-PL}.
}
\end{figure}

The elastic scattering angular distribution at $E_\alpha = 24.7$ MeV by Budzanowski {\it et al.}\ \cite{Budzanowski64-PL} provides the chance to obtain one further data point for \stot\ at the upper end of the energies under study. This value for \stot\ directly constrains the A-OMP at relatively high energies.

The angular distribution of \cite{Budzanowski64-PL} was analyzed in the following way. Optical model fits were performed using either Woods-Saxon potentials of volume type in the real and imaginary part of the OMP, or a folding potential was used in the real part in combination with a surface Woods-Saxon potental in the imaginary part. Furthermore, a phase shift fit was made using the approach of \cite{Chiste96-PRC}. The fits are shown in Fig.~{\ref{fig:e25scat}} and give \stot\ of 563 mb, 525 mb, and 504 mb. Because of the significantly larger $\chi^2$ of the Woods-Saxon fit (which clearly underestimates the elastic cross section around $\vartheta \approx 60^\circ$ and thus slightly overestimates \stot\ with 563 mb), we adopt \stot\ = $520 \pm 20$ mb at 24.7 MeV.

\subsection{$\chi^2$-based assessment}
\label{sec:chi2}
The new experimental data, in combination with the additional data point for \stot\ from \auvii \raa \auvii\ elastic scattering \cite{Budzanowski64-PL} and the further data points of \cite{Basunia07-PRC}, will be used to determine a best-fit set of parameters for the SM calculations. For this purpose the TALYS code (version 1.9) \cite{TALYS-V19} was used which is a well-established open-source code for SM calculations. Similar to previous studies (see \cite{Mohr17-PRC} for $^{64}$Zn + \al , \cite{Talwar18-PRC,Mohr18-PRC} for $^{38}$Ar + \al\ and \cite{Kiss18-PRC} for $^{115}$In + \al ), the complete TALYS parameter space was investigated,  and a $\chi^2$-based assessment was used to find the best description of the experimental data. In practice, 14 different A-OMPs were used which turn out to be the most important ingredient of the SM calculation. These 14 A-OMPs were combined with 5 different N-OMPs, 6 LDs, and 8 GSF (with two options for the M1 contribution), leading to an overall calculation of 6720 excitation functions. The N-OMPs, LDs, and GSFs were taken from the built-in TALYS options. For the A-OMPs 14 different options were used which exceed the 8 standard options in TALYS; these 14 A-OMPs will be discussed in further detail. The A-OMPs are also summarized in Table \ref{tab:aomp}.
%
\begin{table}[t]
\caption{\al -nucleus optical model potentials (A-OMPs): TALYS standard and extensions}
\label{tab:aomp}
\center
\begin{ruledtabular}
\begin{tabular}{cccp{5cm}}
{\it{alphaomp}} & Ref. & Abbr. & comments \\
\noalign{\smallskip}\colrule\noalign{\smallskip}	
1 & \cite{Watanabe58-NP} & WAT
& Watanabe: default in earlier TALYS versions \\
2 & \cite{McFadden66-NP} & MCF
& McFadden/Satchler: simple 4-parameter potential \\
3 & \cite{Demetriou02-NPA} & DEM1
& Demetriou {\it et al.}, version 1: real folding, imaginary volume WS \\
4 & \cite{Demetriou02-NPA} & DEM2
& Demetriou {\it et al.}, version 2: real folding, imaginary volume+surface WS \\
5 & \cite{Demetriou02-NPA} & DEM3
& Demetriou {\it et al.}, version 3: real folding plus dispersion relation \\
6 & \cite{Avrigeanu14-PRC} & AVR
& Avrigeanu {\it et al.}: multi-parameter WS \\
7 & \cite{Nolte87-PRC} & --
& Nolte {\it et al.}: not appropriate for low energies \\
8 & \cite{Avrigeanu94-PRC} & --
& Avrigeanu {\it et al.}: not appropriate for low energies \\
9 & \cite{Mohr13-ADNDT} & AT-V1
& Mohr {\it et al.}: systematic potential, adjusted to low-energy scattering data \\
10 & \cite{Demetriou02-NPA} & DEM3x1.1
& Demetriou {\it et al.}, version 3: real part multiplied by 1.1 \\
11 & \cite{Demetriou02-NPA} & DEM3x1.2
& Demetriou {\it et al.}, version 3: real part multiplied by 1.2 \\
12 & \cite{Demetriou02-NPA} & DEM3x0.9
& Demetriou {\it et al.}, version 3: real part multiplied by 0.9 \\
13 & \cite{Demetriou02-NPA} & DEM3x0.8
& Demetriou {\it et al.}, version 3: real part multiplied by 0.8 \\
14 & \cite{Demetriou02-NPA} & DEM3x0.7
& Demetriou {\it et al.}, version 3: real part multiplied by 0.7 \\
\end{tabular}
\end{ruledtabular}
\end{table}

It is well-known that the early A-OMP by Watanabe (WAT) \cite{Watanabe58-NP}
and the simple 4-parameter Woods-Saxon (WS) potential by McFadden and Satchler
(MCF) \cite{McFadden66-NP} ({\it{alphaomp}} 1 and 2 in TALYS) show a trend to
overestimate the cross sections of \al -induced reactions. This trend becomes
pronounced especially towards low energies below the Coulomb barrier. For
completeness it has to be mentioned that a new explanation for the failure
of the MCF potential at low energies was provided recently in
\cite{Mohr19-IJMPE}.

A series of A-OMPs was suggested by Demetriou {\it et  al.}\ \cite{Demetriou02-NPA} which are based on the double folding procedure in the real part. The first version DEM1 uses a volume WS potential in the imaginary part where the strength is energy-dependent according to a Brown-Rho parametrization \cite{Brown81-NPA}. In the second version DEM2 the imaginary part is composed of a volume WS and a surface WS. The strength of the real parts in DEM1 and DEM2 is taken from the parametrization of real volume integrals $J_R$ from \al -decay data \cite{Mohr00-PRC}. The third version DEM3 uses an imaginary part very close to DEM2 and additionally introduces the coupling between the real and imaginary part by a dispersion relation. Typically, the DEM1, DEM2, and DEM3 potentials ({\it{alphaomp}} 3, 4, 5 in TALYS) predict smaller cross sections than WAT and MCF. Recently it has been pointed out that an excellent reproduction of experimental data can be obtained if the real part of the DEM3 potential is scaled by factors between 1.1 and 1.2 for heavy nuclei \cite{Scholz16-PLB}; a smaller scaling factor of 0.9 was found for $^{64}$Zn \cite{Mohr17-PRC}. Therefore, different scaling factors for the DEM3 potential were also investigated ({\it{alphaomp}} 10-14).

The recent version of the Avrigeanu potentials \cite{Avrigeanu14-PRC} (AVR, {\it{alphaomp}} 6 in TALYS) consists of a real part in WS parametrization which has been chosen close to folding potentials. The imaginary part is composed of WS volume and surface terms with mass- and energy dependent parameters. Similar to the Demetriou potentials, the AVR potential leads to smaller cross sections than WAT and MCF at low energies.

The potential by Nolte {\it et al.}\ \cite{Nolte87-PRC} ({\it{alphaomp}} 7 in TALYS) and the earlier potential by Avrigeanu {\it et al.}\ \cite{Avrigeanu94-PRC} ({\it{alphaomp}} 8) have been adjusted to experimental data at higher energies. It has been found that these potentials are inappropriate at very low energies \cite{Mohr17-PRC,Mohr18-PRC}. This finding is confirmed in the present work where $\chi^2$ per point of above 50,000 (5,800) was found for the Nolte (early Avrigeanu) potential. These huge $\chi^2$ correspond to average deviations from the experimental data by more than a factor of 3.6 (2.4) whereas all other potentials reach average deviations far below a factor of two.

The ATOMKI-V1 potential \cite{Mohr13-ADNDT} (AT-V1, implemented as {\it{alphaomp}} 9 in TALYS V1.8) is based on a double-folding potential in the real part in combination with a surface WS potential in the imaginary part. The parameters of AT-V1 have been adjusted to elastic scattering in the $89 \le A \le 144$ mass range, i.e.\ below the \auvii\ nucleus under study in this work.

The 14 A-OMPs in Table \ref{tab:aomp} were used in a strict $\chi^2$-based assessment. The experimental data show a clear preference for the DEM3 potential (multiplied by 1.1 and 1.2)  and the AVR potential. We find $\chi^2$ per point of about 4.6 (DEM3x1.2), 6.1 (AVR), and 6.2 (DEM3x1.1). This corresponds to an average deviation \fdevbar\ of 1.39, 1.41, and 1.50 for the DEM3x1.2, AVR, and DEM3x1.1 potentials. In the following, all $\chi^2$ will be given per experimental data point. \fdevbar\ is defined by
\begin{equation}
\bar{f}_{\rm{dev}} = \left( \prod_i^N f_{{\rm{dev}},i} \right)^{(1/N)}
\label{eq:fdev}
\end{equation}
and \fdev $_{,i}$ is the larger of the ratios $\sigma_{\rm{calc}}/\sigma_{\rm{exp}}$ or $\sigma_{\rm{exp}}/\sigma_{\rm{calc}}$ for the $i$-th experimental data point.

As pointed out above, the sensitivity to the other ingredients of the SM calculations is relatively minor. About 50 different choices of GSF, N-OMP, and LD in combination with the DEM3x1.2 A-OMP result in a minor increase of $\chi^2$ by less than 1.0 and \fdevbar\ between 1.39 and 1.46.

A strict $\chi^2$ assessment is only valid for statistical uncertainties. Unfortunately, the uncertainties of the present data have a significant contribution from systematic uncertainties (see Table \ref{tab:param} and discussion at the end of Sec.~\ref{sec:exp}). An attempt was made to disentangle the relevance of the statistical and systematic uncertainties. For this attempt we restrict ourselves to our new experimental data from Table \ref{tab:XS}.

\begin{figure}[b]
\includegraphics[width=0.99\columnwidth]{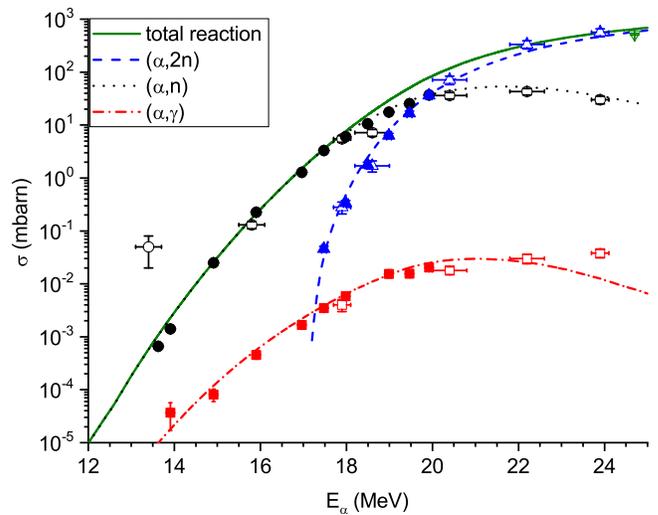}
\caption{\label{fig:XS} Experimental cross sections compared to the best-fit statistical model calculation using the AVR A-OMP. Green down triangle, blue up triangles, black dots, and red squares stands for the total, \rann , \ran , and \rag\ cross sections, respectively. Full symbols are the present data, while open symbols are from \cite{Basunia07-PRC} and the calculated total cross section from \cite{Budzanowski64-PL}. Green full line is the SM predicted total cross sections, while blue dashed, black dotted, and red dot-dashed lines are the \rann , \ran , and \rag\ cross sections, respectively.}
\end{figure}

In a new $\chi^2$ calculation, the best-fit parameters are derived from our experimental data with statistical uncertainties only. Here we find that the best reproduction of our experimental data is obtained for the AVR A-OMP with $\chi^2 = 34.2$ and \fdevbar\ $ = 1.22$. Compared to the result from all available experimental data, we find a significantly increased $\chi^2$ which results from the smaller (statistical only) uncertainties. The average deviation decreases from about 1.4 to 1.2; this decrease is related to relatively large deviation factors \fdev\ for some data points of the Basunia data, in particular at the lowest energy of Basunia for the \ran\ channel. Interestingly, the best-fit A-OMP changes from DEM3x1.2 to AVR; however, the DEM3x1.2 A-OMP provides $\chi^2 \approx 39$ and \fdevbar\ $\approx 1.25$, i.e.\ very close to the results from the AVR A-OMP.
In \fig{fig:XS} the measured cross sections are shown together with the calculated ones using the AVR A-OMP.

Next, we have to consider the systematic uncertainties which are dominated by the \g -ray intensities in the $\beta$-decays of the residual nuclei (see Table \ref{tab:param}). Under these circumstances the systematic uncertainties are common within each \raX\ channel, but not common to all experimental data; i.e., it is possible that both \ran\ and \rann\ data are higher or lower within their systematic uncertainties, but it is also possible that the \ran\ data are higher and the \rann\ data are lower (and vice versa). To cover the full range of systematic uncertainties, we have scaled the \rag , \ran , and \rann\ data by $\pm 2 \sigma$ of the systematic uncertainties, leading to 27 hypothetical experimental data sets within the systematic uncertainties. The number of 27 results from 3 channels which are varied independently by factors $1-2\sigma$, 1.0, and $1+2\sigma$. Further calculations with finer steps in one channel (and no variation in the other channels) confirm that the overall behavior of the $\chi^2$ landscape is relatively smooth.

For the 27 hypothetical data sets, the best-fit parameters of the SM calculations are derived using the same $\chi^2$-based assessment as before. It is found that the best-fit A-OMP is well constrained to the AVR or DEM3x1.2 potentials. In general, the AVR potential is obtained when the \ran\ and \rann\ cross sections are increased whereas the DEM3x1.2 potential is favored for smaller \ran\ and \rann\ cross sections. The overall smallest $\chi^2 = 8.1$ and \fdevbar\ $= 1.18$ is found for the AVR A-OMP and the case where the cross sections of the \rag , \ran , and \rann\ channels are all increased by $2\sigma$ of the systematic uncertainties, i.e., by about 9\% for the \rag , 16\% for the \ran, and 21\% for the \rann\ channel.

Surprisingly, although the present experimental data cover the \rag , \ran ,
and \rann\ channels over several MeV, it is difficult to provide constraints
for the SM parameters beyond the A-OMP. The variation of the experimental data
within their systematic uncertainties constrains the A-OMP to AVR or DEM3x1.2,
but almost any choice of the N-OMP, GSF, and LD appears in the best-fit
parameters of the 27 hypothetical experimental data sets which represent the
range of systematic uncertainties.

\subsection{\label{sec:astro}Extrapolation to astrophysically relevant
  energies}
The A-OMPs are an essential ingredient for the calculation of stellar reaction
rates in the astrophysical $\gamma$ process. This process operates at typical
temperatures of $T_9 \approx 2 - 3$, corresponding to a Gamow window around
8.9 MeV at $T_9 = 2$ and 11.7 MeV at $T_9 = 3$. As the \ran\ channel opens
around 10 MeV and starts to dominate already a few hundred keV above the
threshold, the \rag\ and \ran\ rates above $T_9 \approx 2.5$ must remain
uncertain because the branching between the \rag\ and \ran\ channels depends
on several parameters of the statistical model which cannot be well
constrained (see discussion above). We restrict ourselves to the
analysis of the total reaction cross section \stot\ which depends solely on
the chosen A-OMP.

At $E_{\alpha,{\rm{lab}}} = 11.9$ MeV, corresponding to the center of the
Gamow window at $T_9 \approx 3$, we find \stot\ = 7.34 nb from the AVR
potential and 4.56 nb from the DEM3x1.2 potential; both potentials have been
determined from the $\chi^2$-based assessment in the previous section. Thus,
\stot\ $\approx 6$ nb can be estimated with an uncertainty of about 25\%.
The small uncertainty of 25\% is based on the constraints from the new
experimental data in combination with the similar energy dependence of
\stot\ from the two best-fit potentials down to about 11 MeV.

A further extrapolation down to $E_{\alpha,{\rm{lab}}} = 9.1$ MeV
(corresponding to $T_9 \approx 2$) is obviously more uncertain, but the
predictions from the two best-fit potentials remain within about a factor of
2.5 with \stot\ = 0.123 pb for the AVR potential and \stot\ = 0.047 pb from
the DEM3x1.2 potential. From the average of the two predicted cross sections,
\stot\ $\approx 0.08$ pb with an uncertainty of less than a factor of two can
be recommended. This is a significant achievement, because the range of
predictions at 9.1 MeV from modern A-OMPs (AVR, DEM, AT1) covers two orders of
magnitude from 0.03 to 3 pb, and the MCF potential predicts an even higher
cross section of about 37 pb.

In addition, reaction rates for the \rag\ reaction are calculated at temperatures of $T_9 = 2$ and 3 for the best-fit potentials (DEM3x1.2 and AVR) and compared to the widely used MCF potential. At the higher temperature of $T_9 = 3$ the DEM3x1.2 and AVR rates agree within about a factor of two wheres the MCF rate exceeds the average of the DEM3x1.2 and AVR rates by a factor of 300. The discrepancies increase towards lower temperatures. At $T_9 = 2$ the DEM3x1.2 and AVR rates deviate by a factor of about 3.5, whereas the MCF rate exceeds the DEM3x1.2 and AVR rates by three orders of magnitude.

Finally, it has been pointed out by Rauscher \cite{Rauscher10-PRC} that
the simple Gamow window estimate for the most effective energies is inaccurate
for \al -induced nuclei on heavy nuclei. Typically, the most effective energy
is shifted to lower energies by about $1-2$ MeV, thus further increasing the
range of predicted cross sections at the most effective energies and
increasing the uncertainties of the reaction rates.

\section{\label{sec:sum}Summary}
Alpha-induced reactions were investigated at low energies using the activation
technique in combination with $\gamma$-ray and X-ray spectroscopy. The cross
sections of the \rag , \ran , and \rann\ reactions were measured with
unprecedented sensitivity, and thus far lower cross-section data could be
obtained than available in literature. The lowest data points of the present
work reach the upper end of the Gamow window for temperatures of the
astrophysical $\gamma$ process.

The new dataset allowed us to choose the best $\alpha$-nucleus optical model potential based on a strict
$\chi^2$-based statistical assessment. It was found that the best-fit
theoretical calculations are obtained using either the latest potential by
Avrigeanu {\it et al.}\ \cite{Avrigeanu14-PRC} or the third version of the
Demetriou {\it et al.}\ \cite{Demetriou02-NPA} A-OMP with a scaling factor of
1.2 for the real part. The total reaction cross section is well constrained
within a factor of two uncertainty down to the lowest $\gamma$-process
temperatures. However, due to the systematic uncertainties of the present
data, the other constituents of the statistical model calculations as the
nucleon-nucleus optical model potential, $\gamma$-ray strength function, and level density cannot be constrained; but these constituents typically
have only minor impact for the reaction rate calculations for the most
important \rga\ reactions.

\begin{acknowledgments}
This work was supported by NKFIH (Gr. No. K120666, NN128072), by the New National Excellence Program of the Ministry for Innovation and Technology (\'UNKP-19-3-I-DE-394, \'UNKP-19-4-DE-65), Helmholtz Association (ERC-RA-0016) and European Cooperation in Science and Technology ("ChETEC" COST Action, CA16117).
G.\,G.~Kiss acknowledges support form the J\'anos Bolyai research fellowship of the Hungarian Academy of Sciences.
\end{acknowledgments}

\bibliography{}

\begin{thebibliography}{68}%
\makeatletter
\providecommand \@ifxundefined [1]{%
 \@ifx{#1\undefined}
}%
\providecommand \@ifnum [1]{%
 \ifnum #1\expandafter \@firstoftwo
 \else \expandafter \@secondoftwo
 \fi
}%
\providecommand \@ifx [1]{%
 \ifx #1\expandafter \@firstoftwo
 \else \expandafter \@secondoftwo
 \fi
}%
\providecommand \natexlab [1]{#1}%
\providecommand \enquote  [1]{``#1''}%
\providecommand \bibnamefont  [1]{#1}%
\providecommand \bibfnamefont [1]{#1}%
\providecommand \citenamefont [1]{#1}%
\providecommand \href@noop [0]{\@secondoftwo}%
\providecommand \href [0]{\begingroup \@sanitize@url \@href}%
\providecommand \@href[1]{\@@startlink{#1}\@@href}%
\providecommand \@@href[1]{\endgroup#1\@@endlink}%
\providecommand \@sanitize@url [0]{\catcode `\\12\catcode `\$12\catcode
  `\&12\catcode `\#12\catcode `\^12\catcode `\_12\catcode `\%12\relax}%
\providecommand \@@startlink[1]{}%
\providecommand \@@endlink[0]{}%
\providecommand \url  [0]{\begingroup\@sanitize@url \@url }%
\providecommand \@url [1]{\endgroup\@href {#1}{\urlprefix }}%
\providecommand \urlprefix  [0]{URL }%
\providecommand \Eprint [0]{\href }%
\providecommand \doibase [0]{http://dx.doi.org/}%
\providecommand \selectlanguage [0]{\@gobble}%
\providecommand \bibinfo  [0]{\@secondoftwo}%
\providecommand \bibfield  [0]{\@secondoftwo}%
\providecommand \translation [1]{[#1]}%
\providecommand \BibitemOpen [0]{}%
\providecommand \bibitemStop [0]{}%
\providecommand \bibitemNoStop [0]{.\EOS\space}%
\providecommand \EOS [0]{\spacefactor3000\relax}%
\providecommand \BibitemShut  [1]{\csname bibitem#1\endcsname}%
\let\auto@bib@innerbib\@empty
\bibitem [{\citenamefont {Burbidge}\ \emph {et~al.}(1957)\citenamefont
  {Burbidge}, \citenamefont {Burbidge}, \citenamefont {Fowler},\ and\
  \citenamefont {Hoyle}}]{B2FH57}%
  \BibitemOpen
  \bibfield  {author} {\bibinfo {author} {\bibfnamefont {E.~M.}\ \bibnamefont
  {Burbidge}}, \bibinfo {author} {\bibfnamefont {G.~R.}\ \bibnamefont
  {Burbidge}}, \bibinfo {author} {\bibfnamefont {W.~A.}\ \bibnamefont
  {Fowler}}, \ and\ \bibinfo {author} {\bibfnamefont {F.}~\bibnamefont
  {Hoyle}},\ }\href {\doibase 10.1103/RevModPhys.29.547} {\bibfield  {journal}
  {\bibinfo  {journal} {Rev. Mod. Phys.}\ }\textbf {\bibinfo {volume} {29}},\
  \bibinfo {pages} {547} (\bibinfo {year} {1957})}\BibitemShut {NoStop}%
\bibitem [{\citenamefont {K\"appeler}\ \emph {et~al.}(2011)\citenamefont
  {K\"appeler}, \citenamefont {Gallino}, \citenamefont {Bisterzo},\ and\
  \citenamefont {Aoki}}]{Kappeler11-RMP}%
  \BibitemOpen
  \bibfield  {author} {\bibinfo {author} {\bibfnamefont {F.}~\bibnamefont
  {K\"appeler}}, \bibinfo {author} {\bibfnamefont {R.}~\bibnamefont {Gallino}},
  \bibinfo {author} {\bibfnamefont {S.}~\bibnamefont {Bisterzo}}, \ and\
  \bibinfo {author} {\bibfnamefont {W.}~\bibnamefont {Aoki}},\ }\href {\doibase
  10.1103/RevModPhys.83.157} {\bibfield  {journal} {\bibinfo  {journal} {Rev.
  Mod. Phys.}\ }\textbf {\bibinfo {volume} {83}},\ \bibinfo {pages} {157}
  (\bibinfo {year} {2011})},\ \Eprint {http://arxiv.org/abs/1012.5218}
  {1012.5218} \BibitemShut {NoStop}%
\bibitem [{\citenamefont {Arnould}\ \emph {et~al.}(2007)\citenamefont
  {Arnould}, \citenamefont {Goriely},\ and\ \citenamefont
  {Takahashi}}]{Arnould07-PR}%
  \BibitemOpen
  \bibfield  {author} {\bibinfo {author} {\bibfnamefont {M.}~\bibnamefont
  {Arnould}}, \bibinfo {author} {\bibfnamefont {S.}~\bibnamefont {Goriely}}, \
  and\ \bibinfo {author} {\bibfnamefont {K.}~\bibnamefont {Takahashi}},\ }\href
  {\doibase 10.1016/j.physrep.2007.06.002} {\bibfield  {journal} {\bibinfo
  {journal} {Phys. Rep.}\ }\textbf {\bibinfo {volume} {450}},\ \bibinfo {pages}
  {97} (\bibinfo {year} {2007})},\ \Eprint {http://arxiv.org/abs/0705.4512}
  {0705.4512} \BibitemShut {NoStop}%
\bibitem [{\citenamefont {Rauscher}\ \emph {et~al.}(2013)\citenamefont
  {Rauscher}, \citenamefont {Dauphas}, \citenamefont {Dillmann}, \citenamefont
  {Fr\"ohlich}, \citenamefont {F\"ul\"op},\ and\ \citenamefont
  {Gy\"urky}}]{Rauscher13-RPP}%
  \BibitemOpen
  \bibfield  {author} {\bibinfo {author} {\bibfnamefont {T.}~\bibnamefont
  {Rauscher}}, \bibinfo {author} {\bibfnamefont {N.}~\bibnamefont {Dauphas}},
  \bibinfo {author} {\bibfnamefont {I.}~\bibnamefont {Dillmann}}, \bibinfo
  {author} {\bibfnamefont {C.}~\bibnamefont {Fr\"ohlich}}, \bibinfo {author}
  {\bibfnamefont {Z.}~\bibnamefont {F\"ul\"op}}, \ and\ \bibinfo {author}
  {\bibfnamefont {G.}~\bibnamefont {Gy\"urky}},\ }\href {\doibase
  10.1088/0034-4885/76/6/066201} {\bibfield  {journal} {\bibinfo  {journal}
  {Rep. Prog. Phys.}\ }\textbf {\bibinfo {volume} {76}},\ \bibinfo {pages}
  {066201} (\bibinfo {year} {2013})},\ \Eprint {http://arxiv.org/abs/1303.2666}
  {1303.2666} \BibitemShut {NoStop}%
\bibitem [{\citenamefont {Nishimura}\ \emph {et~al.}(2018)\citenamefont
  {Nishimura}, \citenamefont {Rauscher}, \citenamefont {Hirschi}, \citenamefont
  {Murphy}, \citenamefont {Cescutti},\ and\ \citenamefont
  {Travaglio}}]{Nishimura18-MNRAS}%
  \BibitemOpen
  \bibfield  {author} {\bibinfo {author} {\bibfnamefont {N.}~\bibnamefont
  {Nishimura}}, \bibinfo {author} {\bibfnamefont {T.}~\bibnamefont {Rauscher}},
  \bibinfo {author} {\bibfnamefont {R.}~\bibnamefont {Hirschi}}, \bibinfo
  {author} {\bibfnamefont {A.~S.~J.}\ \bibnamefont {Murphy}}, \bibinfo {author}
  {\bibfnamefont {G.}~\bibnamefont {Cescutti}}, \ and\ \bibinfo {author}
  {\bibfnamefont {C.}~\bibnamefont {Travaglio}},\ }\href {\doibase
  10.1093/mnras/stx3033} {\bibfield  {journal} {\bibinfo  {journal} {Mon. Not.
  R. Astron. Soc.}\ }\textbf {\bibinfo {volume} {474}},\ \bibinfo {pages}
  {3133} (\bibinfo {year} {2018})},\ \Eprint {http://arxiv.org/abs/1711.09098}
  {1711.09098} \BibitemShut {NoStop}%
\bibitem [{\citenamefont {Travaglio}\ \emph {et~al.}(2018)\citenamefont
  {Travaglio}, \citenamefont {Rauscher}, \citenamefont {Heger}, \citenamefont
  {Pignatari},\ and\ \citenamefont {West}}]{Travaglio18-AJ}%
  \BibitemOpen
  \bibfield  {author} {\bibinfo {author} {\bibfnamefont {C.}~\bibnamefont
  {Travaglio}}, \bibinfo {author} {\bibfnamefont {T.}~\bibnamefont {Rauscher}},
  \bibinfo {author} {\bibfnamefont {A.}~\bibnamefont {Heger}}, \bibinfo
  {author} {\bibfnamefont {M.}~\bibnamefont {Pignatari}}, \ and\ \bibinfo
  {author} {\bibfnamefont {C.}~\bibnamefont {West}},\ }\href {\doibase
  10.3847/1538-4357/aaa4f7} {\bibfield  {journal} {\bibinfo  {journal}
  {Astrophys. J.}\ }\textbf {\bibinfo {volume} {854}},\ \bibinfo {pages} {18}
  (\bibinfo {year} {2018})},\ \Eprint {http://arxiv.org/abs/1801.01929}
  {1801.01929} \BibitemShut {NoStop}%
\bibitem [{\citenamefont {Woosley}\ and\ \citenamefont
  {Howard}(1978)}]{Woosley78-AJS}%
  \BibitemOpen
  \bibfield  {author} {\bibinfo {author} {\bibfnamefont {S.~E.}\ \bibnamefont
  {Woosley}}\ and\ \bibinfo {author} {\bibfnamefont {W.~M.}\ \bibnamefont
  {Howard}},\ }\href {\doibase 10.1086/190501} {\bibfield  {journal} {\bibinfo
  {journal} {Astrophys. J. Suppl. S.}\ }\textbf {\bibinfo {volume} {36}},\
  \bibinfo {pages} {285} (\bibinfo {year} {1978})}\BibitemShut {NoStop}%
\bibitem [{\citenamefont {Rauscher}\ \emph {et~al.}(2002)\citenamefont
  {Rauscher}, \citenamefont {Heger}, \citenamefont {Hoffman},\ and\
  \citenamefont {Woosley}}]{Rauscher02-AJ}%
  \BibitemOpen
  \bibfield  {author} {\bibinfo {author} {\bibfnamefont {T.}~\bibnamefont
  {Rauscher}}, \bibinfo {author} {\bibfnamefont {A.}~\bibnamefont {Heger}},
  \bibinfo {author} {\bibfnamefont {R.~D.}\ \bibnamefont {Hoffman}}, \ and\
  \bibinfo {author} {\bibfnamefont {S.~E.}\ \bibnamefont {Woosley}},\ }\href
  {\doibase 10.1086/341728} {\bibfield  {journal} {\bibinfo  {journal}
  {Astrophys. J.}\ }\textbf {\bibinfo {volume} {576}},\ \bibinfo {pages} {323}
  (\bibinfo {year} {2002})},\ \Eprint {http://arxiv.org/abs/astro-ph/0112478}
  {astro-ph/0112478} \BibitemShut {NoStop}%
\bibitem [{\citenamefont {Hauser}\ and\ \citenamefont
  {Feshbach}(1952)}]{Hauser52-PR}%
  \BibitemOpen
  \bibfield  {author} {\bibinfo {author} {\bibfnamefont {W.}~\bibnamefont
  {Hauser}}\ and\ \bibinfo {author} {\bibfnamefont {H.}~\bibnamefont
  {Feshbach}},\ }\href {\doibase 10.1103/PhysRev.87.366} {\bibfield  {journal}
  {\bibinfo  {journal} {Phys. Rev.}\ }\textbf {\bibinfo {volume} {87}},\
  \bibinfo {pages} {366} (\bibinfo {year} {1952})}\BibitemShut {NoStop}%
\bibitem [{\citenamefont {Kiss}\ \emph {et~al.}(2013)\citenamefont {Kiss},
  \citenamefont {Mohr}, \citenamefont {F\"ul\"op}, \citenamefont {Rauscher},
  \citenamefont {Gy\"urky}, \citenamefont {Sz\"ucs}, \citenamefont
  {Hal{\'{a}}sz}, \citenamefont {Somorjai}, \citenamefont {Ornelas},
  \citenamefont {Yal{\c{c}}{\i}n}, \citenamefont {G\"uray},\ and\ \citenamefont
  {\"Ozkan}}]{Kiss13-PRC}%
  \BibitemOpen
  \bibfield  {author} {\bibinfo {author} {\bibfnamefont {G.~G.}\ \bibnamefont
  {Kiss}}, \bibinfo {author} {\bibfnamefont {P.}~\bibnamefont {Mohr}}, \bibinfo
  {author} {\bibfnamefont {Z.}~\bibnamefont {F\"ul\"op}}, \bibinfo {author}
  {\bibfnamefont {T.}~\bibnamefont {Rauscher}}, \bibinfo {author}
  {\bibfnamefont {G.}~\bibnamefont {Gy\"urky}}, \bibinfo {author}
  {\bibfnamefont {T.}~\bibnamefont {Sz\"ucs}}, \bibinfo {author} {\bibfnamefont
  {Z.}~\bibnamefont {Hal{\'{a}}sz}}, \bibinfo {author} {\bibfnamefont
  {E.}~\bibnamefont {Somorjai}}, \bibinfo {author} {\bibfnamefont
  {A.}~\bibnamefont {Ornelas}}, \bibinfo {author} {\bibfnamefont
  {C.}~\bibnamefont {Yal{\c{c}}{\i}n}}, \bibinfo {author} {\bibfnamefont
  {R.~T.}\ \bibnamefont {G\"uray}}, \ and\ \bibinfo {author} {\bibfnamefont
  {N.}~\bibnamefont {\"Ozkan}},\ }\href {\doibase 10.1103/PhysRevC.88.045804}
  {\bibfield  {journal} {\bibinfo  {journal} {Phys. Rev. C}\ }\textbf {\bibinfo
  {volume} {88}},\ \bibinfo {pages} {045804} (\bibinfo {year} {2013})},\
  \Eprint {http://arxiv.org/abs/1311.0368} {1311.0368} \BibitemShut {NoStop}%
\bibitem [{\citenamefont {Netterdon}\ \emph {et~al.}(2013)\citenamefont
  {Netterdon}, \citenamefont {Demetriou}, \citenamefont {Endres}, \citenamefont
  {Giesen}, \citenamefont {Kiss}, \citenamefont {Sauerwein}, \citenamefont
  {Sz\"ucs}, \citenamefont {Zell},\ and\ \citenamefont
  {Zilges}}]{Netterdon13-NPA}%
  \BibitemOpen
  \bibfield  {author} {\bibinfo {author} {\bibfnamefont {L.}~\bibnamefont
  {Netterdon}}, \bibinfo {author} {\bibfnamefont {P.}~\bibnamefont
  {Demetriou}}, \bibinfo {author} {\bibfnamefont {J.}~\bibnamefont {Endres}},
  \bibinfo {author} {\bibfnamefont {U.}~\bibnamefont {Giesen}}, \bibinfo
  {author} {\bibfnamefont {G.}~\bibnamefont {Kiss}}, \bibinfo {author}
  {\bibfnamefont {A.}~\bibnamefont {Sauerwein}}, \bibinfo {author}
  {\bibfnamefont {T.}~\bibnamefont {Sz\"ucs}}, \bibinfo {author} {\bibfnamefont
  {K.}~\bibnamefont {Zell}}, \ and\ \bibinfo {author} {\bibfnamefont
  {A.}~\bibnamefont {Zilges}},\ }\href {\doibase
  10.1016/j.nuclphysa.2013.08.002} {\bibfield  {journal} {\bibinfo  {journal}
  {Nucl. Phys. A}\ }\textbf {\bibinfo {volume} {916}},\ \bibinfo {pages} {149}
  (\bibinfo {year} {2013})},\ \Eprint {http://arxiv.org/abs/1310.1825}
  {1310.1825} \BibitemShut {NoStop}%
\bibitem [{\citenamefont {\"Ozkan}\ \emph {et~al.}(2007)\citenamefont
  {\"Ozkan}, \citenamefont {Efe}, \citenamefont {G\"uray}, \citenamefont
  {Palumbo}, \citenamefont {G\"orres}, \citenamefont {Lee}, \citenamefont
  {Lamm}, \citenamefont {Rapp}, \citenamefont {Stech}, \citenamefont
  {Wiescher}, \citenamefont {Gy\"urky}, \citenamefont {F\"ul\"op},\ and\
  \citenamefont {Somorjai}}]{Ozkan07-PRC}%
  \BibitemOpen
  \bibfield  {author} {\bibinfo {author} {\bibfnamefont {N.}~\bibnamefont
  {\"Ozkan}}, \bibinfo {author} {\bibfnamefont {G.}~\bibnamefont {Efe}},
  \bibinfo {author} {\bibfnamefont {R.~T.}\ \bibnamefont {G\"uray}}, \bibinfo
  {author} {\bibfnamefont {A.}~\bibnamefont {Palumbo}}, \bibinfo {author}
  {\bibfnamefont {J.}~\bibnamefont {G\"orres}}, \bibinfo {author}
  {\bibfnamefont {H.~Y.}\ \bibnamefont {Lee}}, \bibinfo {author} {\bibfnamefont
  {L.~O.}\ \bibnamefont {Lamm}}, \bibinfo {author} {\bibfnamefont
  {W.}~\bibnamefont {Rapp}}, \bibinfo {author} {\bibfnamefont {E.}~\bibnamefont
  {Stech}}, \bibinfo {author} {\bibfnamefont {M.}~\bibnamefont {Wiescher}},
  \bibinfo {author} {\bibfnamefont {G.}~\bibnamefont {Gy\"urky}}, \bibinfo
  {author} {\bibfnamefont {Z.}~\bibnamefont {F\"ul\"op}}, \ and\ \bibinfo
  {author} {\bibfnamefont {E.}~\bibnamefont {Somorjai}},\ }\href {\doibase
  10.1103/physrevc.75.025801} {\bibfield  {journal} {\bibinfo  {journal} {Phys.
  Rev. C}\ }\textbf {\bibinfo {volume} {75}},\ \bibinfo {pages} {025801}
  (\bibinfo {year} {2007})}\BibitemShut {NoStop}%
\bibitem [{\citenamefont {Sauerwein}\ \emph {et~al.}(2011)\citenamefont
  {Sauerwein}, \citenamefont {Becker}, \citenamefont {Dombrowski},
  \citenamefont {Elvers}, \citenamefont {Endres}, \citenamefont {Giesen},
  \citenamefont {Hasper}, \citenamefont {Hennig}, \citenamefont {Netterdon},
  \citenamefont {Rauscher}, \citenamefont {Rogalla}, \citenamefont {Zell},\
  and\ \citenamefont {Zilges}}]{Sauerwein11-PRC}%
  \BibitemOpen
  \bibfield  {author} {\bibinfo {author} {\bibfnamefont {A.}~\bibnamefont
  {Sauerwein}}, \bibinfo {author} {\bibfnamefont {H.-W.}\ \bibnamefont
  {Becker}}, \bibinfo {author} {\bibfnamefont {H.}~\bibnamefont {Dombrowski}},
  \bibinfo {author} {\bibfnamefont {M.}~\bibnamefont {Elvers}}, \bibinfo
  {author} {\bibfnamefont {J.}~\bibnamefont {Endres}}, \bibinfo {author}
  {\bibfnamefont {U.}~\bibnamefont {Giesen}}, \bibinfo {author} {\bibfnamefont
  {J.}~\bibnamefont {Hasper}}, \bibinfo {author} {\bibfnamefont
  {A.}~\bibnamefont {Hennig}}, \bibinfo {author} {\bibfnamefont
  {L.}~\bibnamefont {Netterdon}}, \bibinfo {author} {\bibfnamefont
  {T.}~\bibnamefont {Rauscher}}, \bibinfo {author} {\bibfnamefont
  {D.}~\bibnamefont {Rogalla}}, \bibinfo {author} {\bibfnamefont {K.~O.}\
  \bibnamefont {Zell}}, \ and\ \bibinfo {author} {\bibfnamefont
  {A.}~\bibnamefont {Zilges}},\ }\href {\doibase 10.1103/PhysRevC.84.045808}
  {\bibfield  {journal} {\bibinfo  {journal} {Phys. Rev. C}\ }\textbf {\bibinfo
  {volume} {84}},\ \bibinfo {pages} {045808} (\bibinfo {year} {2011})},\
  \Eprint {http://arxiv.org/abs/1110.4789} {1110.4789} \BibitemShut {NoStop}%
\bibitem [{\citenamefont {Filipescu}\ \emph {et~al.}(2011)\citenamefont
  {Filipescu}, \citenamefont {Avrigeanu}, \citenamefont {Glodariu},
  \citenamefont {Mihai}, \citenamefont {Bucurescu}, \citenamefont
  {Iva{\c{s}}cu}, \citenamefont {C{\u{a}}ta-Danil}, \citenamefont {Stroe},
  \citenamefont {Sima}, \citenamefont {C{\u{a}}ta-Danil}, \citenamefont
  {Deleanu}, \citenamefont {Ghi{\c{t}}{\u{a}}}, \citenamefont
  {M{\u{a}}rginean}, \citenamefont {M{\u{a}}rginean}, \citenamefont {Negret},
  \citenamefont {Pascu}, \citenamefont {Sava}, \citenamefont {Suliman},\ and\
  \citenamefont {Zamfir}}]{Filipescu11-PRC}%
  \BibitemOpen
  \bibfield  {author} {\bibinfo {author} {\bibfnamefont {D.}~\bibnamefont
  {Filipescu}}, \bibinfo {author} {\bibfnamefont {V.}~\bibnamefont
  {Avrigeanu}}, \bibinfo {author} {\bibfnamefont {T.}~\bibnamefont {Glodariu}},
  \bibinfo {author} {\bibfnamefont {C.}~\bibnamefont {Mihai}}, \bibinfo
  {author} {\bibfnamefont {D.}~\bibnamefont {Bucurescu}}, \bibinfo {author}
  {\bibfnamefont {M.}~\bibnamefont {Iva{\c{s}}cu}}, \bibinfo {author}
  {\bibfnamefont {I.}~\bibnamefont {C{\u{a}}ta-Danil}}, \bibinfo {author}
  {\bibfnamefont {L.}~\bibnamefont {Stroe}}, \bibinfo {author} {\bibfnamefont
  {O.}~\bibnamefont {Sima}}, \bibinfo {author} {\bibfnamefont {G.}~\bibnamefont
  {C{\u{a}}ta-Danil}}, \bibinfo {author} {\bibfnamefont {D.}~\bibnamefont
  {Deleanu}}, \bibinfo {author} {\bibfnamefont {D.~G.}\ \bibnamefont
  {Ghi{\c{t}}{\u{a}}}}, \bibinfo {author} {\bibfnamefont {N.}~\bibnamefont
  {M{\u{a}}rginean}}, \bibinfo {author} {\bibfnamefont {R.}~\bibnamefont
  {M{\u{a}}rginean}}, \bibinfo {author} {\bibfnamefont {A.}~\bibnamefont
  {Negret}}, \bibinfo {author} {\bibfnamefont {S.}~\bibnamefont {Pascu}},
  \bibinfo {author} {\bibfnamefont {T.}~\bibnamefont {Sava}}, \bibinfo {author}
  {\bibfnamefont {G.}~\bibnamefont {Suliman}}, \ and\ \bibinfo {author}
  {\bibfnamefont {N.~V.}\ \bibnamefont {Zamfir}},\ }\href {\doibase
  10.1103/PhysRevC.83.064609} {\bibfield  {journal} {\bibinfo  {journal} {Phys.
  Rev. C}\ }\textbf {\bibinfo {volume} {83}},\ \bibinfo {pages} {064609}
  (\bibinfo {year} {2011})}\BibitemShut {NoStop}%
\bibitem [{\citenamefont {Kiss}\ \emph {et~al.}(2014)\citenamefont {Kiss},
  \citenamefont {Sz\"ucs}, \citenamefont {Rauscher}, \citenamefont {T\"or\"ok},
  \citenamefont {F\"ul\"op}, \citenamefont {Gy\"urky}, \citenamefont
  {Hal{\'{a}}sz},\ and\ \citenamefont {Somorjai}}]{Kiss14-PLB}%
  \BibitemOpen
  \bibfield  {author} {\bibinfo {author} {\bibfnamefont {G.~G.}\ \bibnamefont
  {Kiss}}, \bibinfo {author} {\bibfnamefont {T.}~\bibnamefont {Sz\"ucs}},
  \bibinfo {author} {\bibfnamefont {T.}~\bibnamefont {Rauscher}}, \bibinfo
  {author} {\bibfnamefont {Z.}~\bibnamefont {T\"or\"ok}}, \bibinfo {author}
  {\bibfnamefont {Z.}~\bibnamefont {F\"ul\"op}}, \bibinfo {author}
  {\bibfnamefont {G.}~\bibnamefont {Gy\"urky}}, \bibinfo {author}
  {\bibfnamefont {Z.}~\bibnamefont {Hal{\'{a}}sz}}, \ and\ \bibinfo {author}
  {\bibfnamefont {E.}~\bibnamefont {Somorjai}},\ }\href {\doibase
  10.1016/j.physletb.2014.06.011} {\bibfield  {journal} {\bibinfo  {journal}
  {Phys. Lett. B}\ }\textbf {\bibinfo {volume} {735}},\ \bibinfo {pages} {40}
  (\bibinfo {year} {2014})},\ \Eprint {http://arxiv.org/abs/1406.4989}
  {1406.4989} \BibitemShut {NoStop}%
\bibitem [{\citenamefont {Simon}\ \emph {et~al.}(2015)\citenamefont {Simon},
  \citenamefont {Beard}, \citenamefont {Spyrou}, \citenamefont {Quinn},
  \citenamefont {Bucher}, \citenamefont {Couder}, \citenamefont {DeYoung},
  \citenamefont {Dombos}, \citenamefont {G\"orres}, \citenamefont {Kontos},
  \citenamefont {Long}, \citenamefont {Moran}, \citenamefont {Paul},
  \citenamefont {Pereira}, \citenamefont {Robertson}, \citenamefont {Smith},
  \citenamefont {Stech}, \citenamefont {Talwar}, \citenamefont {Tan},\ and\
  \citenamefont {Wiescher}}]{Simon15-PRC}%
  \BibitemOpen
  \bibfield  {author} {\bibinfo {author} {\bibfnamefont {A.}~\bibnamefont
  {Simon}}, \bibinfo {author} {\bibfnamefont {M.}~\bibnamefont {Beard}},
  \bibinfo {author} {\bibfnamefont {A.}~\bibnamefont {Spyrou}}, \bibinfo
  {author} {\bibfnamefont {S.~J.}\ \bibnamefont {Quinn}}, \bibinfo {author}
  {\bibfnamefont {B.}~\bibnamefont {Bucher}}, \bibinfo {author} {\bibfnamefont
  {M.}~\bibnamefont {Couder}}, \bibinfo {author} {\bibfnamefont {P.~A.}\
  \bibnamefont {DeYoung}}, \bibinfo {author} {\bibfnamefont {A.~C.}\
  \bibnamefont {Dombos}}, \bibinfo {author} {\bibfnamefont {J.}~\bibnamefont
  {G\"orres}}, \bibinfo {author} {\bibfnamefont {A.}~\bibnamefont {Kontos}},
  \bibinfo {author} {\bibfnamefont {A.}~\bibnamefont {Long}}, \bibinfo {author}
  {\bibfnamefont {M.~T.}\ \bibnamefont {Moran}}, \bibinfo {author}
  {\bibfnamefont {N.}~\bibnamefont {Paul}}, \bibinfo {author} {\bibfnamefont
  {J.}~\bibnamefont {Pereira}}, \bibinfo {author} {\bibfnamefont
  {D.}~\bibnamefont {Robertson}}, \bibinfo {author} {\bibfnamefont
  {K.}~\bibnamefont {Smith}}, \bibinfo {author} {\bibfnamefont
  {E.}~\bibnamefont {Stech}}, \bibinfo {author} {\bibfnamefont
  {R.}~\bibnamefont {Talwar}}, \bibinfo {author} {\bibfnamefont {W.~P.}\
  \bibnamefont {Tan}}, \ and\ \bibinfo {author} {\bibfnamefont
  {M.}~\bibnamefont {Wiescher}},\ }\href {\doibase 10.1103/PhysRevC.92.025806}
  {\bibfield  {journal} {\bibinfo  {journal} {Phys. Rev. C}\ }\textbf {\bibinfo
  {volume} {92}},\ \bibinfo {pages} {025806} (\bibinfo {year}
  {2015})}\BibitemShut {NoStop}%
\bibitem [{\citenamefont {Yal{\c{c}}{\i}n}\ \emph {et~al.}(2015)\citenamefont
  {Yal{\c{c}}{\i}n}, \citenamefont {Gy\"urky}, \citenamefont {Rauscher},
  \citenamefont {Kiss}, \citenamefont {\"Ozkan}, \citenamefont {G\"uray},
  \citenamefont {Hal\'asz}, \citenamefont {Sz\"ucs}, \citenamefont {F\"ul\"op},
  \citenamefont {Farkas}, \citenamefont {Korkulu},\ and\ \citenamefont
  {Somorjai}}]{Yalcin15-PRC}%
  \BibitemOpen
  \bibfield  {author} {\bibinfo {author} {\bibfnamefont {C.}~\bibnamefont
  {Yal{\c{c}}{\i}n}}, \bibinfo {author} {\bibfnamefont {G.}~\bibnamefont
  {Gy\"urky}}, \bibinfo {author} {\bibfnamefont {T.}~\bibnamefont {Rauscher}},
  \bibinfo {author} {\bibfnamefont {G.~G.}\ \bibnamefont {Kiss}}, \bibinfo
  {author} {\bibfnamefont {N.}~\bibnamefont {\"Ozkan}}, \bibinfo {author}
  {\bibfnamefont {R.~T.}\ \bibnamefont {G\"uray}}, \bibinfo {author}
  {\bibfnamefont {Z.}~\bibnamefont {Hal\'asz}}, \bibinfo {author}
  {\bibfnamefont {T.}~\bibnamefont {Sz\"ucs}}, \bibinfo {author} {\bibfnamefont
  {Z.}~\bibnamefont {F\"ul\"op}}, \bibinfo {author} {\bibfnamefont
  {J.}~\bibnamefont {Farkas}}, \bibinfo {author} {\bibfnamefont
  {Z.}~\bibnamefont {Korkulu}}, \ and\ \bibinfo {author} {\bibfnamefont
  {E.}~\bibnamefont {Somorjai}},\ }\href {\doibase 10.1103/PhysRevC.91.034610}
  {\bibfield  {journal} {\bibinfo  {journal} {Phys. Rev. C}\ }\textbf {\bibinfo
  {volume} {91}},\ \bibinfo {pages} {034610} (\bibinfo {year} {2015})},\
  \Eprint {http://arxiv.org/abs/1504.01651} {1504.01651} \BibitemShut {NoStop}%
\bibitem [{\citenamefont {Hal{\'{a}}sz}\ \emph {et~al.}(2016)\citenamefont
  {Hal{\'{a}}sz}, \citenamefont {Somorjai}, \citenamefont {Gy\"urky},
  \citenamefont {Elekes}, \citenamefont {F\"ul\"op}, \citenamefont {Sz\"ucs},
  \citenamefont {Kiss}, \citenamefont {Szegedi}, \citenamefont {Rauscher},
  \citenamefont {G\"orres},\ and\ \citenamefont {Wiescher}}]{Halasz16-PRC}%
  \BibitemOpen
  \bibfield  {author} {\bibinfo {author} {\bibfnamefont {Z.}~\bibnamefont
  {Hal{\'{a}}sz}}, \bibinfo {author} {\bibfnamefont {E.}~\bibnamefont
  {Somorjai}}, \bibinfo {author} {\bibfnamefont {G.}~\bibnamefont {Gy\"urky}},
  \bibinfo {author} {\bibfnamefont {Z.}~\bibnamefont {Elekes}}, \bibinfo
  {author} {\bibfnamefont {Z.}~\bibnamefont {F\"ul\"op}}, \bibinfo {author}
  {\bibfnamefont {T.}~\bibnamefont {Sz\"ucs}}, \bibinfo {author} {\bibfnamefont
  {G.~G.}\ \bibnamefont {Kiss}}, \bibinfo {author} {\bibfnamefont {N.~T.}\
  \bibnamefont {Szegedi}}, \bibinfo {author} {\bibfnamefont {T.}~\bibnamefont
  {Rauscher}}, \bibinfo {author} {\bibfnamefont {J.}~\bibnamefont {G\"orres}},
  \ and\ \bibinfo {author} {\bibfnamefont {M.}~\bibnamefont {Wiescher}},\
  }\href {\doibase 10.1103/PhysRevC.94.045801} {\bibfield  {journal} {\bibinfo
  {journal} {Phys. Rev. C}\ }\textbf {\bibinfo {volume} {94}},\ \bibinfo
  {pages} {045801} (\bibinfo {year} {2016})},\ \Eprint
  {http://arxiv.org/abs/1609.05612} {1609.05612} \BibitemShut {NoStop}%
\bibitem [{\citenamefont {Scholz}\ \emph {et~al.}(2016)\citenamefont {Scholz},
  \citenamefont {Heim}, \citenamefont {Mayer}, \citenamefont {M\"uker},
  \citenamefont {Netterdon}, \citenamefont {Wombacher},\ and\ \citenamefont
  {Zilges}}]{Scholz16-PLB}%
  \BibitemOpen
  \bibfield  {author} {\bibinfo {author} {\bibfnamefont {P.}~\bibnamefont
  {Scholz}}, \bibinfo {author} {\bibfnamefont {F.}~\bibnamefont {Heim}},
  \bibinfo {author} {\bibfnamefont {J.}~\bibnamefont {Mayer}}, \bibinfo
  {author} {\bibfnamefont {C.}~\bibnamefont {M\"uker}}, \bibinfo {author}
  {\bibfnamefont {L.}~\bibnamefont {Netterdon}}, \bibinfo {author}
  {\bibfnamefont {F.}~\bibnamefont {Wombacher}}, \ and\ \bibinfo {author}
  {\bibfnamefont {A.}~\bibnamefont {Zilges}},\ }\href {\doibase
  10.1016/j.physletb.2016.08.040} {\bibfield  {journal} {\bibinfo  {journal}
  {Phys. Lett. B}\ }\textbf {\bibinfo {volume} {761}},\ \bibinfo {pages} {247}
  (\bibinfo {year} {2016})}\BibitemShut {NoStop}%
\bibitem [{\citenamefont {Mayer}\ \emph {et~al.}(2016)\citenamefont {Mayer},
  \citenamefont {Goriely}, \citenamefont {Netterdon}, \citenamefont {P\'eru},
  \citenamefont {Scholz}, \citenamefont {Schwengner},\ and\ \citenamefont
  {Zilges}}]{Mayer16-PRC}%
  \BibitemOpen
  \bibfield  {author} {\bibinfo {author} {\bibfnamefont {J.}~\bibnamefont
  {Mayer}}, \bibinfo {author} {\bibfnamefont {S.}~\bibnamefont {Goriely}},
  \bibinfo {author} {\bibfnamefont {L.}~\bibnamefont {Netterdon}}, \bibinfo
  {author} {\bibfnamefont {S.}~\bibnamefont {P\'eru}}, \bibinfo {author}
  {\bibfnamefont {P.}~\bibnamefont {Scholz}}, \bibinfo {author} {\bibfnamefont
  {R.}~\bibnamefont {Schwengner}}, \ and\ \bibinfo {author} {\bibfnamefont
  {A.}~\bibnamefont {Zilges}},\ }\href {\doibase 10.1103/PhysRevC.93.045809}
  {\bibfield  {journal} {\bibinfo  {journal} {Phys. Rev. C}\ }\textbf {\bibinfo
  {volume} {93}},\ \bibinfo {pages} {045809} (\bibinfo {year}
  {2016})}\BibitemShut {NoStop}%
\bibitem [{\citenamefont {Sz\"ucs}\ \emph {et~al.}(2018)\citenamefont
  {Sz\"ucs}, \citenamefont {Kiss}, \citenamefont {Gy\"urky}, \citenamefont
  {Hal{\'{a}}sz}, \citenamefont {F\"ul\"op},\ and\ \citenamefont
  {Rauscher}}]{Szucs18-PLB}%
  \BibitemOpen
  \bibfield  {author} {\bibinfo {author} {\bibfnamefont {T.}~\bibnamefont
  {Sz\"ucs}}, \bibinfo {author} {\bibfnamefont {G.}~\bibnamefont {Kiss}},
  \bibinfo {author} {\bibfnamefont {G.}~\bibnamefont {Gy\"urky}}, \bibinfo
  {author} {\bibfnamefont {Z.}~\bibnamefont {Hal{\'{a}}sz}}, \bibinfo {author}
  {\bibfnamefont {Z.}~\bibnamefont {F\"ul\"op}}, \ and\ \bibinfo {author}
  {\bibfnamefont {T.}~\bibnamefont {Rauscher}},\ }\href {\doibase
  10.1016/j.physletb.2017.11.072} {\bibfield  {journal} {\bibinfo  {journal}
  {Phys. Lett. B}\ }\textbf {\bibinfo {volume} {776}},\ \bibinfo {pages} {396}
  (\bibinfo {year} {2018})},\ \Eprint {http://arxiv.org/abs/1711.11271}
  {1711.11271} \BibitemShut {NoStop}%
\bibitem [{\citenamefont {Korkulu}\ \emph {et~al.}(2018)\citenamefont
  {Korkulu}, \citenamefont {\"Ozkan}, \citenamefont {Kiss}, \citenamefont
  {Sz\"ucs}, \citenamefont {Gy\"urky}, \citenamefont {F\"ul\"op}, \citenamefont
  {G\"uray}, \citenamefont {Hal{\'{a}}sz}, \citenamefont {Rauscher},
  \citenamefont {Somorjai}, \citenamefont {T\"or\"ok},\ and\ \citenamefont
  {Yal{\c{c}}{\i}n}}]{Korkulu18-PRC}%
  \BibitemOpen
  \bibfield  {author} {\bibinfo {author} {\bibfnamefont {Z.}~\bibnamefont
  {Korkulu}}, \bibinfo {author} {\bibfnamefont {N.}~\bibnamefont {\"Ozkan}},
  \bibinfo {author} {\bibfnamefont {G.~G.}\ \bibnamefont {Kiss}}, \bibinfo
  {author} {\bibfnamefont {T.}~\bibnamefont {Sz\"ucs}}, \bibinfo {author}
  {\bibfnamefont {G.}~\bibnamefont {Gy\"urky}}, \bibinfo {author}
  {\bibfnamefont {Z.}~\bibnamefont {F\"ul\"op}}, \bibinfo {author}
  {\bibfnamefont {R.~T.}\ \bibnamefont {G\"uray}}, \bibinfo {author}
  {\bibfnamefont {Z.}~\bibnamefont {Hal{\'{a}}sz}}, \bibinfo {author}
  {\bibfnamefont {T.}~\bibnamefont {Rauscher}}, \bibinfo {author}
  {\bibfnamefont {E.}~\bibnamefont {Somorjai}}, \bibinfo {author}
  {\bibfnamefont {Z.}~\bibnamefont {T\"or\"ok}}, \ and\ \bibinfo {author}
  {\bibfnamefont {C.}~\bibnamefont {Yal{\c{c}}{\i}n}},\ }\href {\doibase
  10.1103/PhysRevC.97.045803} {\bibfield  {journal} {\bibinfo  {journal} {Phys.
  Rev. C}\ }\textbf {\bibinfo {volume} {97}},\ \bibinfo {pages} {045803}
  (\bibinfo {year} {2018})},\ \Eprint {http://arxiv.org/abs/1803.07791}
  {1803.07791} \BibitemShut {NoStop}%
\bibitem [{\citenamefont {Kiss}\ \emph {et~al.}(2018)\citenamefont {Kiss},
  \citenamefont {Sz\"ucs}, \citenamefont {Mohr}, \citenamefont {T\"or\"ok},
  \citenamefont {Husz{\'{a}}nk}, \citenamefont {Gy\"urky},\ and\ \citenamefont
  {F\"ul\"op}}]{Kiss18-PRC}%
  \BibitemOpen
  \bibfield  {author} {\bibinfo {author} {\bibfnamefont {G.~G.}\ \bibnamefont
  {Kiss}}, \bibinfo {author} {\bibfnamefont {T.}~\bibnamefont {Sz\"ucs}},
  \bibinfo {author} {\bibfnamefont {P.}~\bibnamefont {Mohr}}, \bibinfo {author}
  {\bibfnamefont {Z.}~\bibnamefont {T\"or\"ok}}, \bibinfo {author}
  {\bibfnamefont {R.}~\bibnamefont {Husz{\'{a}}nk}}, \bibinfo {author}
  {\bibfnamefont {G.}~\bibnamefont {Gy\"urky}}, \ and\ \bibinfo {author}
  {\bibfnamefont {Z.}~\bibnamefont {F\"ul\"op}},\ }\href {\doibase
  10.1103/physrevc.97.055803} {\bibfield  {journal} {\bibinfo  {journal} {Phys.
  Rev. C}\ }\textbf {\bibinfo {volume} {97}},\ \bibinfo {pages} {055803}
  (\bibinfo {year} {2018})},\ \Eprint {http://arxiv.org/abs/1805.07533}
  {1805.07533} \BibitemShut {NoStop}%
\bibitem [{\citenamefont {Gy\"urky}\ \emph {et~al.}(2019)\citenamefont
  {Gy\"urky}, \citenamefont {F\"ul\"op}, \citenamefont {K\"appeler},
  \citenamefont {Kiss},\ and\ \citenamefont {Wallner}}]{Gyurky19-EPJA}%
  \BibitemOpen
  \bibfield  {author} {\bibinfo {author} {\bibfnamefont {G.}~\bibnamefont
  {Gy\"urky}}, \bibinfo {author} {\bibfnamefont {Z.}~\bibnamefont {F\"ul\"op}},
  \bibinfo {author} {\bibfnamefont {F.}~\bibnamefont {K\"appeler}}, \bibinfo
  {author} {\bibfnamefont {G.~G.}\ \bibnamefont {Kiss}}, \ and\ \bibinfo
  {author} {\bibfnamefont {A.}~\bibnamefont {Wallner}},\ }\href {\doibase
  10.1140/epja/i2019-12708-4} {\bibfield  {journal} {\bibinfo  {journal} {Eur.
  Phys. J. A}\ }\textbf {\bibinfo {volume} {55}},\ \bibinfo {pages} {41}
  (\bibinfo {year} {2019})},\ \Eprint {http://arxiv.org/abs/1903.03339}
  {1903.03339} \BibitemShut {NoStop}%
\bibitem [{\citenamefont {Singh}(2007)}]{NDS199}%
  \BibitemOpen
  \bibfield  {author} {\bibinfo {author} {\bibfnamefont {B.}~\bibnamefont
  {Singh}},\ }\href {\doibase 10.1016/j.nds.2007.01.001} {\bibfield  {journal}
  {\bibinfo  {journal} {Nucl. Data Sheets}\ }\textbf {\bibinfo {volume}
  {108}},\ \bibinfo {pages} {79} (\bibinfo {year} {2007})}\BibitemShut
  {NoStop}%
\bibitem [{\citenamefont {Kondev}\ and\ \citenamefont
  {Lalkovski}(2007)}]{NDS200}%
  \BibitemOpen
  \bibfield  {author} {\bibinfo {author} {\bibfnamefont {F.~G.}\ \bibnamefont
  {Kondev}}\ and\ \bibinfo {author} {\bibfnamefont {S.}~\bibnamefont
  {Lalkovski}},\ }\href {\doibase 10.1016/j.nds.2007.06.002} {\bibfield
  {journal} {\bibinfo  {journal} {Nucl. Data Sheets}\ }\textbf {\bibinfo
  {volume} {108}},\ \bibinfo {pages} {1471} (\bibinfo {year}
  {2007})}\BibitemShut {NoStop}%
\bibitem [{\citenamefont {Kondev}(2007)}]{NDS201}%
  \BibitemOpen
  \bibfield  {author} {\bibinfo {author} {\bibfnamefont {F.~G.}\ \bibnamefont
  {Kondev}},\ }\href {\doibase 10.1016/j.nds.2007.01.004} {\bibfield  {journal}
  {\bibinfo  {journal} {Nucl. Data Sheets}\ }\textbf {\bibinfo {volume}
  {108}},\ \bibinfo {pages} {365} (\bibinfo {year} {2007})}\BibitemShut
  {NoStop}%
\bibitem [{\citenamefont {Basunia}\ \emph {et~al.}(2007)\citenamefont
  {Basunia}, \citenamefont {Shugart}, \citenamefont {Smith},\ and\
  \citenamefont {Norman}}]{Basunia07-PRC}%
  \BibitemOpen
  \bibfield  {author} {\bibinfo {author} {\bibfnamefont {M.~S.}\ \bibnamefont
  {Basunia}}, \bibinfo {author} {\bibfnamefont {H.~A.}\ \bibnamefont
  {Shugart}}, \bibinfo {author} {\bibfnamefont {A.~R.}\ \bibnamefont {Smith}},
  \ and\ \bibinfo {author} {\bibfnamefont {E.~B.}\ \bibnamefont {Norman}},\
  }\href {\doibase 10.1103/PhysRevC.75.015802} {\bibfield  {journal} {\bibinfo
  {journal} {Phys. Rev. C}\ }\textbf {\bibinfo {volume} {75}},\ \bibinfo
  {pages} {015802} (\bibinfo {year} {2007})}\BibitemShut {NoStop}%
\bibitem [{\citenamefont {Capurro}\ \emph {et~al.}(1988)\citenamefont
  {Capurro}, \citenamefont {de~la Vega~Vedoya},\ and\ \citenamefont
  {Nassiff}}]{Capurro88-JRNC}%
  \BibitemOpen
  \bibfield  {author} {\bibinfo {author} {\bibfnamefont {O.~A.}\ \bibnamefont
  {Capurro}}, \bibinfo {author} {\bibfnamefont {M.}~\bibnamefont {de~la
  Vega~Vedoya}}, \ and\ \bibinfo {author} {\bibfnamefont {S.~J.}\ \bibnamefont
  {Nassiff}},\ }\href {\doibase 10.1007/BF02205195} {\bibfield  {journal}
  {\bibinfo  {journal} {J. Radioanal. Nucl. Chem.}\ }\textbf {\bibinfo {volume}
  {128}},\ \bibinfo {pages} {403} (\bibinfo {year} {1988})}\BibitemShut
  {NoStop}%
\bibitem [{\citenamefont {Necheva}\ and\ \citenamefont
  {Kolev}(1997)}]{Necheva97-ARI}%
  \BibitemOpen
  \bibfield  {author} {\bibinfo {author} {\bibfnamefont {C.}~\bibnamefont
  {Necheva}}\ and\ \bibinfo {author} {\bibfnamefont {D.}~\bibnamefont
  {Kolev}},\ }\href {\doibase 10.1016/s0969-8043(96)00277-1} {\bibfield
  {journal} {\bibinfo  {journal} {Appl. Radiat. Isot.}\ }\textbf {\bibinfo
  {volume} {48}},\ \bibinfo {pages} {807} (\bibinfo {year} {1997})}\BibitemShut
  {NoStop}%
\bibitem [{\citenamefont {Kulko}\ \emph {et~al.}(2007)\citenamefont {Kulko},
  \citenamefont {Demekhina}, \citenamefont {Kalpakchieva}, \citenamefont
  {Muzychka}, \citenamefont {Penionzhkevich}, \citenamefont {Rassadov},
  \citenamefont {Skobelev},\ and\ \citenamefont {Testov}}]{Kulko07-PAN}%
  \BibitemOpen
  \bibfield  {author} {\bibinfo {author} {\bibfnamefont {A.~A.}\ \bibnamefont
  {Kulko}}, \bibinfo {author} {\bibfnamefont {N.~A.}\ \bibnamefont
  {Demekhina}}, \bibinfo {author} {\bibfnamefont {R.}~\bibnamefont
  {Kalpakchieva}}, \bibinfo {author} {\bibfnamefont {Y.~A.}\ \bibnamefont
  {Muzychka}}, \bibinfo {author} {\bibfnamefont {Y.~E.}\ \bibnamefont
  {Penionzhkevich}}, \bibinfo {author} {\bibfnamefont {D.~N.}\ \bibnamefont
  {Rassadov}}, \bibinfo {author} {\bibfnamefont {N.~K.}\ \bibnamefont
  {Skobelev}}, \ and\ \bibinfo {author} {\bibfnamefont {D.~A.}\ \bibnamefont
  {Testov}},\ }\href {\doibase 10.1134/s1063778807040011} {\bibfield  {journal}
  {\bibinfo  {journal} {Phys. At. Nucl.}\ }\textbf {\bibinfo {volume} {70}},\
  \bibinfo {pages} {613} (\bibinfo {year} {2007})}\BibitemShut {NoStop}%
\bibitem [{\citenamefont {Kurz}\ \emph {et~al.}(1971)\citenamefont {Kurz},
  \citenamefont {Jasper}, \citenamefont {Fischer},\ and\ \citenamefont
  {Hermes}}]{Kurz71-NPA}%
  \BibitemOpen
  \bibfield  {author} {\bibinfo {author} {\bibfnamefont {H.}~\bibnamefont
  {Kurz}}, \bibinfo {author} {\bibfnamefont {E.}~\bibnamefont {Jasper}},
  \bibinfo {author} {\bibfnamefont {K.}~\bibnamefont {Fischer}}, \ and\
  \bibinfo {author} {\bibfnamefont {F.}~\bibnamefont {Hermes}},\ }\href
  {\doibase 10.1016/0375-9474(71)90654-3} {\bibfield  {journal} {\bibinfo
  {journal} {Nucl. Phys. A}\ }\textbf {\bibinfo {volume} {168}},\ \bibinfo
  {pages} {129} (\bibinfo {year} {1971})}\BibitemShut {NoStop}%
\bibitem [{\citenamefont {Calboreanu}\ \emph {et~al.}(1982)\citenamefont
  {Calboreanu}, \citenamefont {Pencea},\ and\ \citenamefont
  {Salagean}}]{Calboreanu82-NPA}%
  \BibitemOpen
  \bibfield  {author} {\bibinfo {author} {\bibfnamefont {A.}~\bibnamefont
  {Calboreanu}}, \bibinfo {author} {\bibfnamefont {C.}~\bibnamefont {Pencea}},
  \ and\ \bibinfo {author} {\bibfnamefont {O.}~\bibnamefont {Salagean}},\
  }\href {\doibase 10.1016/0375-9474(82)90451-1} {\bibfield  {journal}
  {\bibinfo  {journal} {Nucl. Phys. A}\ }\textbf {\bibinfo {volume} {383}},\
  \bibinfo {pages} {251} (\bibinfo {year} {1982})}\BibitemShut {NoStop}%
\bibitem [{\citenamefont {Capurro}\ \emph {et~al.}(1985)\citenamefont
  {Capurro}, \citenamefont {de~la Vega~Vedoya}, \citenamefont {Wasilevsky},\
  and\ \citenamefont {Nassiff}}]{Capurro85-JRNC}%
  \BibitemOpen
  \bibfield  {author} {\bibinfo {author} {\bibfnamefont {O.~A.}\ \bibnamefont
  {Capurro}}, \bibinfo {author} {\bibfnamefont {M.}~\bibnamefont {de~la
  Vega~Vedoya}}, \bibinfo {author} {\bibfnamefont {C.}~\bibnamefont
  {Wasilevsky}}, \ and\ \bibinfo {author} {\bibfnamefont {S.~J.}\ \bibnamefont
  {Nassiff}},\ }\href {\doibase 10.1007/bf02040615} {\bibfield  {journal}
  {\bibinfo  {journal} {Journal of Radioanalytical and Nuclear Chemistry}\
  }\textbf {\bibinfo {volume} {89}},\ \bibinfo {pages} {519} (\bibinfo {year}
  {1985})}\BibitemShut {NoStop}%
\bibitem [{\citenamefont {Bhardwaj}\ and\ \citenamefont
  {Prasad}(1986)}]{Bhardwaj86-NIMA}%
  \BibitemOpen
  \bibfield  {author} {\bibinfo {author} {\bibfnamefont {H.}~\bibnamefont
  {Bhardwaj}}\ and\ \bibinfo {author} {\bibfnamefont {R.}~\bibnamefont
  {Prasad}},\ }\href {\doibase 10.1016/0168-9002(86)90222-6} {\bibfield
  {journal} {\bibinfo  {journal} {Nucl. Instrum. Meth. A}\ }\textbf {\bibinfo
  {volume} {242}},\ \bibinfo {pages} {286} (\bibinfo {year}
  {1986})}\BibitemShut {NoStop}%
\bibitem [{\citenamefont {Shah}\ \emph {et~al.}(1995)\citenamefont {Shah},
  \citenamefont {Patel}, \citenamefont {Singh}, \citenamefont {Mukherjee},\
  and\ \citenamefont {Chintalapudi}}]{Shah95-Pra}%
  \BibitemOpen
  \bibfield  {author} {\bibinfo {author} {\bibfnamefont {D.~J.}\ \bibnamefont
  {Shah}}, \bibinfo {author} {\bibfnamefont {H.~B.}\ \bibnamefont {Patel}},
  \bibinfo {author} {\bibfnamefont {N.~L.}\ \bibnamefont {Singh}}, \bibinfo
  {author} {\bibfnamefont {S.}~\bibnamefont {Mukherjee}}, \ and\ \bibinfo
  {author} {\bibfnamefont {S.~N.}\ \bibnamefont {Chintalapudi}},\ }\href
  {\doibase 10.1007/bf02878864} {\bibfield  {journal} {\bibinfo  {journal}
  {Pramana}\ }\textbf {\bibinfo {volume} {44}},\ \bibinfo {pages} {535}
  (\bibinfo {year} {1995})}\BibitemShut {NoStop}%
\bibitem [{\citenamefont {Ismail}(1998)}]{Ismail98-Pra}%
  \BibitemOpen
  \bibfield  {author} {\bibinfo {author} {\bibfnamefont {M.}~\bibnamefont
  {Ismail}},\ }\href {\doibase 10.1007/BF02847528} {\bibfield  {journal}
  {\bibinfo  {journal} {Pramana}\ }\textbf {\bibinfo {volume} {50}},\ \bibinfo
  {pages} {173} (\bibinfo {year} {1998})}\BibitemShut {NoStop}%
\bibitem [{\citenamefont {Lanzafame}\ and\ \citenamefont
  {Blann}(1970)}]{Lanzafame70-NPA}%
  \BibitemOpen
  \bibfield  {author} {\bibinfo {author} {\bibfnamefont {F.}~\bibnamefont
  {Lanzafame}}\ and\ \bibinfo {author} {\bibfnamefont {M.}~\bibnamefont
  {Blann}},\ }\href {\doibase 10.1016/0375-9474(70)90811-0} {\bibfield
  {journal} {\bibinfo  {journal} {Nucl. Phys. A}\ }\textbf {\bibinfo {volume}
  {142}},\ \bibinfo {pages} {545} (\bibinfo {year} {1970})}\BibitemShut
  {NoStop}%
\bibitem [{\citenamefont {Koltay}\ \emph {et~al.}(2011)\citenamefont {Koltay},
  \citenamefont {P{\'{a}}szti},\ and\ \citenamefont {Kiss}}]{Koltay11-HNC}%
  \BibitemOpen
  \bibfield  {author} {\bibinfo {author} {\bibfnamefont {E.}~\bibnamefont
  {Koltay}}, \bibinfo {author} {\bibfnamefont {F.}~\bibnamefont
  {P{\'{a}}szti}}, \ and\ \bibinfo {author} {\bibfnamefont {{\'{A}}.~Z.}\
  \bibnamefont {Kiss}},\ }in\ \href {\doibase 10.1007/978-1-4419-0720-2_33}
  {\emph {\bibinfo {booktitle} {Handbook of Nuclear Chemistry}}},\ \bibinfo
  {editor} {edited by\ \bibinfo {editor} {\bibfnamefont {A.}~\bibnamefont
  {V\'ertes}}, \bibinfo {editor} {\bibfnamefont {S.}~\bibnamefont {Nagy}},
  \bibinfo {editor} {\bibfnamefont {Z.}~\bibnamefont {Klencs\'ar}}, \bibinfo
  {editor} {\bibfnamefont {R.~G.}\ \bibnamefont {Lovas}}, \ and\ \bibinfo
  {editor} {\bibfnamefont {F.}~\bibnamefont {R\"osch}}}\ (\bibinfo  {publisher}
  {Springer {US}},\ \bibinfo {year} {2011})\ pp.\ \bibinfo {pages}
  {1695--1735}\BibitemShut {NoStop}%
\bibitem [{\citenamefont {Kert{\'{e}}sz}\ \emph {et~al.}(2010)\citenamefont
  {Kert{\'{e}}sz}, \citenamefont {Szoboszlai}, \citenamefont {Angyal},
  \citenamefont {Dobos},\ and\ \citenamefont
  {Borb{\'{e}}ly-Kiss}}]{Kertesz10-NIMB}%
  \BibitemOpen
  \bibfield  {author} {\bibinfo {author} {\bibfnamefont {Z.}~\bibnamefont
  {Kert{\'{e}}sz}}, \bibinfo {author} {\bibfnamefont {Z.}~\bibnamefont
  {Szoboszlai}}, \bibinfo {author} {\bibfnamefont {A.}~\bibnamefont {Angyal}},
  \bibinfo {author} {\bibfnamefont {E.}~\bibnamefont {Dobos}}, \ and\ \bibinfo
  {author} {\bibfnamefont {I.}~\bibnamefont {Borb{\'{e}}ly-Kiss}},\ }\href
  {\doibase 10.1016/j.nimb.2010.02.103} {\bibfield  {journal} {\bibinfo
  {journal} {Nucl. Instrum. Meth. B}\ }\textbf {\bibinfo {volume} {268}},\
  \bibinfo {pages} {1924} (\bibinfo {year} {2010})}\BibitemShut {NoStop}%
\bibitem [{\citenamefont {Campbell}\ \emph {et~al.}(2010)\citenamefont
  {Campbell}, \citenamefont {Boyd}, \citenamefont {Grassi}, \citenamefont
  {Bonnick},\ and\ \citenamefont {Maxwell}}]{Campbell10-NIMB}%
  \BibitemOpen
  \bibfield  {author} {\bibinfo {author} {\bibfnamefont {J.}~\bibnamefont
  {Campbell}}, \bibinfo {author} {\bibfnamefont {N.}~\bibnamefont {Boyd}},
  \bibinfo {author} {\bibfnamefont {N.}~\bibnamefont {Grassi}}, \bibinfo
  {author} {\bibfnamefont {P.}~\bibnamefont {Bonnick}}, \ and\ \bibinfo
  {author} {\bibfnamefont {J.}~\bibnamefont {Maxwell}},\ }\href {\doibase
  10.1016/j.nimb.2010.07.012} {\bibfield  {journal} {\bibinfo  {journal} {Nucl.
  Instrum. Meth. B}\ }\textbf {\bibinfo {volume} {268}},\ \bibinfo {pages}
  {3356} (\bibinfo {year} {2010})}\BibitemShut {NoStop}%
\bibitem [{\citenamefont {Husz\'ank}\ \emph {et~al.}(2015)\citenamefont
  {Husz\'ank}, \citenamefont {Csedreki}, \citenamefont {Kert{\'{e}}sz},\ and\
  \citenamefont {T\"or\"ok}}]{Huszank15-JRNC}%
  \BibitemOpen
  \bibfield  {author} {\bibinfo {author} {\bibfnamefont {R.}~\bibnamefont
  {Husz\'ank}}, \bibinfo {author} {\bibfnamefont {L.}~\bibnamefont {Csedreki}},
  \bibinfo {author} {\bibfnamefont {Z.}~\bibnamefont {Kert{\'{e}}sz}}, \ and\
  \bibinfo {author} {\bibfnamefont {Z.}~\bibnamefont {T\"or\"ok}},\ }\href
  {\doibase 10.1007/s10967-015-4102-9} {\bibfield  {journal} {\bibinfo
  {journal} {J. Radioanal. Nucl. Chem.}\ }\textbf {\bibinfo {volume} {307}},\
  \bibinfo {pages} {341} (\bibinfo {year} {2015})}\BibitemShut {NoStop}%
\bibitem [{\citenamefont {Mayer}(2011)}]{SIMNRA}%
  \BibitemOpen
  \bibfield  {author} {\bibinfo {author} {\bibfnamefont {M.}~\bibnamefont
  {Mayer}},\ }\href {https://home.mpcdf.mpg.de/~mam/} {} (\bibinfo {year}
  {2011}),\ \bibinfo {note} {{SIMNRA} version 6.06,
  https://home.mpcdf.mpg.de/~mam/}\BibitemShut {NoStop}%
\bibitem [{\citenamefont {Sz\"ucs}\ \emph {et~al.}(2014)\citenamefont
  {Sz\"ucs}, \citenamefont {Kiss},\ and\ \citenamefont
  {F\"ul\"op}}]{Szucs14-AIPConf}%
  \BibitemOpen
  \bibfield  {author} {\bibinfo {author} {\bibfnamefont {T.}~\bibnamefont
  {Sz\"ucs}}, \bibinfo {author} {\bibfnamefont {G.~G.}\ \bibnamefont {Kiss}}, \
  and\ \bibinfo {author} {\bibfnamefont {Z.}~\bibnamefont {F\"ul\"op}},\ }\href
  {\doibase 10.1063/1.4875306} {\bibfield  {journal} {\bibinfo  {journal} {AIP
  Conf. Proc.}\ }\textbf {\bibinfo {volume} {1595}},\ \bibinfo {pages} {173}
  (\bibinfo {year} {2014})}\BibitemShut {NoStop}%
\bibitem [{\citenamefont {Ziegler}\ \emph {et~al.}(2010)\citenamefont
  {Ziegler}, \citenamefont {Ziegler},\ and\ \citenamefont {Biersack}}]{srim}%
  \BibitemOpen
  \bibfield  {author} {\bibinfo {author} {\bibfnamefont {J.~F.}\ \bibnamefont
  {Ziegler}}, \bibinfo {author} {\bibfnamefont {M.}~\bibnamefont {Ziegler}}, \
  and\ \bibinfo {author} {\bibfnamefont {J.}~\bibnamefont {Biersack}},\ }\href
  {\doibase 10.1016/j.nimb.2010.02.091} {\bibfield  {journal} {\bibinfo
  {journal} {Nucl. Instrum. Meth. B}\ }\textbf {\bibinfo {volume} {268}},\
  \bibinfo {pages} {1818} (\bibinfo {year} {2010})},\ \bibinfo {note}
  {http://www.srim.org}\BibitemShut {NoStop}%
\bibitem [{\citenamefont {Debertin}\ \emph {et~al.}(1979)\citenamefont
  {Debertin}, \citenamefont {Pessara}, \citenamefont {Sch\"otzig},\ and\
  \citenamefont {Walz}}]{Debertin79-ARI}%
  \BibitemOpen
  \bibfield  {author} {\bibinfo {author} {\bibfnamefont {K.}~\bibnamefont
  {Debertin}}, \bibinfo {author} {\bibfnamefont {W.}~\bibnamefont {Pessara}},
  \bibinfo {author} {\bibfnamefont {U.}~\bibnamefont {Sch\"otzig}}, \ and\
  \bibinfo {author} {\bibfnamefont {K.}~\bibnamefont {Walz}},\ }\href {\doibase
  10.1016/0020-708x(79)90169-8} {\bibfield  {journal} {\bibinfo  {journal}
  {Int. J. Appl. Radiat. Isot.}\ }\textbf {\bibinfo {volume} {30}},\ \bibinfo
  {pages} {551} (\bibinfo {year} {1979})}\BibitemShut {NoStop}%
\bibitem [{\citenamefont {Kinsey}\ \emph {et~al.}(1997)\citenamefont {Kinsey}
  \emph {et~al.}}]{NuDat}%
  \BibitemOpen
  \bibfield  {author} {\bibinfo {author} {\bibfnamefont {R.~R.}\ \bibnamefont
  {Kinsey}} \emph {et~al.},\ }in\ \href@noop {} {\emph {\bibinfo {booktitle}
  {Proceedings of the 9th International Symposium on Capture gamma-ray
  spectroscopy and related topics.}}},\ \bibinfo {editor} {edited by\ \bibinfo
  {editor} {\bibfnamefont {G.~L.}\ \bibnamefont {Molnar}}, \bibinfo {editor}
  {\bibfnamefont {T.}~\bibnamefont {Belgya}}, \ and\ \bibinfo {editor}
  {\bibfnamefont {Z.}~\bibnamefont {Revay}}}\ (\bibinfo  {publisher} {Springer;
  Budapest (Hungary)},\ \bibinfo {year} {1997})\ p.\ \bibinfo {pages} {506},\
  \bibinfo {note} {data extracted from the {NuDat 2.7} database (26. 06.
  2019)}\BibitemShut {NoStop}%
\bibitem [{\citenamefont {Burrows}(1988)}]{RADLIST}%
  \BibitemOpen
  \bibfield  {author} {\bibinfo {author} {\bibfnamefont {T.~W.}\ \bibnamefont
  {Burrows}},\ }\href@noop {} {\emph {\bibinfo {title} {The Program
  {RADLIST}}}},\ \bibinfo {type} {Tech. Rep.}\ (\bibinfo  {institution} {Report
  BNL-NCS-52142},\ \bibinfo {year} {1988})\BibitemShut {NoStop}%
\bibitem [{\citenamefont {Rauscher}(2011)}]{Rauscher11-IJMPE}%
  \BibitemOpen
  \bibfield  {author} {\bibinfo {author} {\bibfnamefont {T.}~\bibnamefont
  {Rauscher}},\ }\href {\doibase 10.1142/s021830131101840x} {\bibfield
  {journal} {\bibinfo  {journal} {Int. J Mod. Phys. E}\ }\textbf {\bibinfo
  {volume} {20}},\ \bibinfo {pages} {1071} (\bibinfo {year} {2011})},\ \Eprint
  {http://arxiv.org/abs/1010.4283} {1010.4283} \BibitemShut {NoStop}%
\bibitem [{\citenamefont {Gy\"urky}\ \emph {et~al.}(2012)\citenamefont
  {Gy\"urky}, \citenamefont {Mohr}, \citenamefont {F\"ul\"op}, \citenamefont
  {Hal\'asz}, \citenamefont {Kiss}, \citenamefont {Sz\"ucs},\ and\
  \citenamefont {Somorjai}}]{Gyurky12-PRC}%
  \BibitemOpen
  \bibfield  {author} {\bibinfo {author} {\bibfnamefont {G.}~\bibnamefont
  {Gy\"urky}}, \bibinfo {author} {\bibfnamefont {P.}~\bibnamefont {Mohr}},
  \bibinfo {author} {\bibfnamefont {Z.}~\bibnamefont {F\"ul\"op}}, \bibinfo
  {author} {\bibfnamefont {Z.}~\bibnamefont {Hal\'asz}}, \bibinfo {author}
  {\bibfnamefont {G.~G.}\ \bibnamefont {Kiss}}, \bibinfo {author}
  {\bibfnamefont {T.}~\bibnamefont {Sz\"ucs}}, \ and\ \bibinfo {author}
  {\bibfnamefont {E.}~\bibnamefont {Somorjai}},\ }\href {\doibase
  10.1103/physrevc.86.041601} {\bibfield  {journal} {\bibinfo  {journal} {Phys.
  Rev. C}\ }\textbf {\bibinfo {volume} {86}},\ \bibinfo {pages} {041601(R)}
  (\bibinfo {year} {2012})}\BibitemShut {NoStop}%
\bibitem [{\citenamefont {Ornelas}\ \emph {et~al.}(2016)\citenamefont
  {Ornelas}, \citenamefont {Mohr}, \citenamefont {Gy\"urky}, \citenamefont
  {Elekes}, \citenamefont {F\"ul\"op}, \citenamefont {Hal{\'{a}}sz},
  \citenamefont {Kiss}, \citenamefont {Somorjai}, \citenamefont {Sz\"ucs},
  \citenamefont {Tak\'{a}cs}, \citenamefont {Galaviz}, \citenamefont {G\"uray},
  \citenamefont {Korkulu}, \citenamefont {\"Ozkan},\ and\ \citenamefont
  {Yal{\c{c}}{\i}n}}]{Ornelas16-PRC}%
  \BibitemOpen
  \bibfield  {author} {\bibinfo {author} {\bibfnamefont {A.}~\bibnamefont
  {Ornelas}}, \bibinfo {author} {\bibfnamefont {P.}~\bibnamefont {Mohr}},
  \bibinfo {author} {\bibfnamefont {G.}~\bibnamefont {Gy\"urky}}, \bibinfo
  {author} {\bibfnamefont {Z.}~\bibnamefont {Elekes}}, \bibinfo {author}
  {\bibfnamefont {Z.}~\bibnamefont {F\"ul\"op}}, \bibinfo {author}
  {\bibfnamefont {Z.}~\bibnamefont {Hal{\'{a}}sz}}, \bibinfo {author}
  {\bibfnamefont {G.~G.}\ \bibnamefont {Kiss}}, \bibinfo {author}
  {\bibfnamefont {E.}~\bibnamefont {Somorjai}}, \bibinfo {author}
  {\bibfnamefont {T.}~\bibnamefont {Sz\"ucs}}, \bibinfo {author} {\bibfnamefont
  {M.~P.}\ \bibnamefont {Tak\'{a}cs}}, \bibinfo {author} {\bibfnamefont
  {D.}~\bibnamefont {Galaviz}}, \bibinfo {author} {\bibfnamefont {R.~T.}\
  \bibnamefont {G\"uray}}, \bibinfo {author} {\bibfnamefont {Z.}~\bibnamefont
  {Korkulu}}, \bibinfo {author} {\bibfnamefont {N.}~\bibnamefont {\"Ozkan}}, \
  and\ \bibinfo {author} {\bibfnamefont {C.}~\bibnamefont {Yal{\c{c}}{\i}n}},\
  }\href {\doibase 10.1103/PhysRevC.94.055807} {\bibfield  {journal} {\bibinfo
  {journal} {Phys. Rev. C}\ }\textbf {\bibinfo {volume} {94}},\ \bibinfo
  {pages} {055807} (\bibinfo {year} {2016})},\ \Eprint
  {http://arxiv.org/abs/1611.00927} {1611.00927} \BibitemShut {NoStop}%
\bibitem [{\citenamefont {Budzanowski}\ \emph {et~al.}(1964)\citenamefont
  {Budzanowski}, \citenamefont {Grotowski}, \citenamefont {Micek},
  \citenamefont {Niewodnicza{\'{n}}ski}, \citenamefont {{\'{S}}li{\r{z}}},
  \citenamefont {Strza{\l}kowski},\ and\ \citenamefont
  {Wojciechowski}}]{Budzanowski64-PL}%
  \BibitemOpen
  \bibfield  {author} {\bibinfo {author} {\bibfnamefont {A.}~\bibnamefont
  {Budzanowski}}, \bibinfo {author} {\bibfnamefont {K.}~\bibnamefont
  {Grotowski}}, \bibinfo {author} {\bibfnamefont {S.}~\bibnamefont {Micek}},
  \bibinfo {author} {\bibfnamefont {H.}~\bibnamefont {Niewodnicza{\'{n}}ski}},
  \bibinfo {author} {\bibfnamefont {J.}~\bibnamefont {{\'{S}}li{\r{z}}}},
  \bibinfo {author} {\bibfnamefont {A.}~\bibnamefont {Strza{\l}kowski}}, \ and\
  \bibinfo {author} {\bibfnamefont {H.}~\bibnamefont {Wojciechowski}},\ }\href
  {\doibase 10.1016/0031-9163(64)90265-3} {\bibfield  {journal} {\bibinfo
  {journal} {Phys. Lett.}\ }\textbf {\bibinfo {volume} {11}},\ \bibinfo {pages}
  {74} (\bibinfo {year} {1964})}\BibitemShut {NoStop}%
\bibitem [{\citenamefont {Chist\'e}\ \emph {et~al.}(1996)\citenamefont
  {Chist\'e}, \citenamefont {Lichtenth\"aler}, \citenamefont {Villari},\ and\
  \citenamefont {Gomes}}]{Chiste96-PRC}%
  \BibitemOpen
  \bibfield  {author} {\bibinfo {author} {\bibfnamefont {V.}~\bibnamefont
  {Chist\'e}}, \bibinfo {author} {\bibfnamefont {R.}~\bibnamefont
  {Lichtenth\"aler}}, \bibinfo {author} {\bibfnamefont {A.~C.~C.}\ \bibnamefont
  {Villari}}, \ and\ \bibinfo {author} {\bibfnamefont {L.~C.}\ \bibnamefont
  {Gomes}},\ }\href {\doibase 10.1103/PhysRevC.54.784} {\bibfield  {journal}
  {\bibinfo  {journal} {Phys. Rev. C}\ }\textbf {\bibinfo {volume} {54}},\
  \bibinfo {pages} {784} (\bibinfo {year} {1996})}\BibitemShut {NoStop}%
\bibitem [{\citenamefont {Koning}\ \emph {et~al.}(2017)\citenamefont {Koning},
  \citenamefont {Hilaire},\ and\ \citenamefont {Goriely}}]{TALYS-V19}%
  \BibitemOpen
  \bibfield  {author} {\bibinfo {author} {\bibfnamefont {A.~J.}\ \bibnamefont
  {Koning}}, \bibinfo {author} {\bibfnamefont {S.}~\bibnamefont {Hilaire}}, \
  and\ \bibinfo {author} {\bibfnamefont {S.}~\bibnamefont {Goriely}},\ }\href
  {http://www.talys.eu/} {\enquote {\bibinfo {title} {computer code
  {\sc{talys}}, version 1.9},}\ } (\bibinfo {year} {2017})\BibitemShut
  {NoStop}%
\bibitem [{\citenamefont {Mohr}\ \emph {et~al.}(2017)\citenamefont {Mohr},
  \citenamefont {Gy\"urky},\ and\ \citenamefont {F\"ul\"op}}]{Mohr17-PRC}%
  \BibitemOpen
  \bibfield  {author} {\bibinfo {author} {\bibfnamefont {P.}~\bibnamefont
  {Mohr}}, \bibinfo {author} {\bibfnamefont {G.}~\bibnamefont {Gy\"urky}}, \
  and\ \bibinfo {author} {\bibfnamefont {Z.}~\bibnamefont {F\"ul\"op}},\ }\href
  {\doibase 10.1103/PhysRevC.95.015807} {\bibfield  {journal} {\bibinfo
  {journal} {Phys. Rev. C}\ }\textbf {\bibinfo {volume} {95}},\ \bibinfo
  {pages} {015807} (\bibinfo {year} {2017})},\ \Eprint
  {http://arxiv.org/abs/1701.02290} {1701.02290} \BibitemShut {NoStop}%
\bibitem [{\citenamefont {Talwar}\ \emph {et~al.}(2018)\citenamefont {Talwar},
  \citenamefont {Bojazi}, \citenamefont {Mohr}, \citenamefont {Auranen},
  \citenamefont {Avila}, \citenamefont {Ayangeakaa}, \citenamefont {Harker},
  \citenamefont {Hoffman}, \citenamefont {Jiang}, \citenamefont {Kuvin},
  \citenamefont {Meyer}, \citenamefont {Rehm}, \citenamefont
  {Santiago-Gonzalez}, \citenamefont {Sethi}, \citenamefont {Ugalde},\ and\
  \citenamefont {Winkelbauer}}]{Talwar18-PRC}%
  \BibitemOpen
  \bibfield  {author} {\bibinfo {author} {\bibfnamefont {R.}~\bibnamefont
  {Talwar}}, \bibinfo {author} {\bibfnamefont {M.~J.}\ \bibnamefont {Bojazi}},
  \bibinfo {author} {\bibfnamefont {P.}~\bibnamefont {Mohr}}, \bibinfo {author}
  {\bibfnamefont {K.}~\bibnamefont {Auranen}}, \bibinfo {author} {\bibfnamefont
  {M.~L.}\ \bibnamefont {Avila}}, \bibinfo {author} {\bibfnamefont {A.~D.}\
  \bibnamefont {Ayangeakaa}}, \bibinfo {author} {\bibfnamefont
  {J.}~\bibnamefont {Harker}}, \bibinfo {author} {\bibfnamefont {C.~R.}\
  \bibnamefont {Hoffman}}, \bibinfo {author} {\bibfnamefont {C.~L.}\
  \bibnamefont {Jiang}}, \bibinfo {author} {\bibfnamefont {S.~A.}\ \bibnamefont
  {Kuvin}}, \bibinfo {author} {\bibfnamefont {B.~S.}\ \bibnamefont {Meyer}},
  \bibinfo {author} {\bibfnamefont {K.~E.}\ \bibnamefont {Rehm}}, \bibinfo
  {author} {\bibfnamefont {D.}~\bibnamefont {Santiago-Gonzalez}}, \bibinfo
  {author} {\bibfnamefont {J.}~\bibnamefont {Sethi}}, \bibinfo {author}
  {\bibfnamefont {C.}~\bibnamefont {Ugalde}}, \ and\ \bibinfo {author}
  {\bibfnamefont {J.~R.}\ \bibnamefont {Winkelbauer}},\ }\href {\doibase
  10.1103/PhysRevC.97.055801} {\bibfield  {journal} {\bibinfo  {journal} {Phys.
  Rev. C}\ }\textbf {\bibinfo {volume} {97}},\ \bibinfo {pages} {055801}
  (\bibinfo {year} {2018})}\BibitemShut {NoStop}%
\bibitem [{\citenamefont {Mohr}\ \emph {et~al.}(2018)\citenamefont {Mohr},
  \citenamefont {Talwar},\ and\ \citenamefont {Avila}}]{Mohr18-PRC}%
  \BibitemOpen
  \bibfield  {author} {\bibinfo {author} {\bibfnamefont {P.}~\bibnamefont
  {Mohr}}, \bibinfo {author} {\bibfnamefont {R.}~\bibnamefont {Talwar}}, \ and\
  \bibinfo {author} {\bibfnamefont {M.~L.}\ \bibnamefont {Avila}},\ }\href
  {\doibase 10.1103/PhysRevC.98.045805} {\bibfield  {journal} {\bibinfo
  {journal} {Phys. Rev. C}\ }\textbf {\bibinfo {volume} {98}},\ \bibinfo
  {pages} {045805} (\bibinfo {year} {2018})},\ \Eprint
  {http://arxiv.org/abs/1810.02527} {1810.02527} \BibitemShut {NoStop}%
\bibitem [{\citenamefont {Watanabe}(1958)}]{Watanabe58-NP}%
  \BibitemOpen
  \bibfield  {author} {\bibinfo {author} {\bibfnamefont {S.}~\bibnamefont
  {Watanabe}},\ }\href {\doibase 10.1016/0029-5582(58)90180-9} {\bibfield
  {journal} {\bibinfo  {journal} {Nuclear Physics}\ }\textbf {\bibinfo {volume}
  {8}},\ \bibinfo {pages} {484} (\bibinfo {year} {1958})}\BibitemShut {NoStop}%
\bibitem [{\citenamefont {McFadden}\ and\ \citenamefont
  {Satchler}(1966)}]{McFadden66-NP}%
  \BibitemOpen
  \bibfield  {author} {\bibinfo {author} {\bibfnamefont {L.}~\bibnamefont
  {McFadden}}\ and\ \bibinfo {author} {\bibfnamefont {G.}~\bibnamefont
  {Satchler}},\ }\href {\doibase 10.1016/0029-5582(66)90441-x} {\bibfield
  {journal} {\bibinfo  {journal} {Nuclear Physics}\ }\textbf {\bibinfo {volume}
  {84}},\ \bibinfo {pages} {177} (\bibinfo {year} {1966})}\BibitemShut
  {NoStop}%
\bibitem [{\citenamefont {Demetriou}\ \emph {et~al.}(2002)\citenamefont
  {Demetriou}, \citenamefont {Grama},\ and\ \citenamefont
  {Goriely}}]{Demetriou02-NPA}%
  \BibitemOpen
  \bibfield  {author} {\bibinfo {author} {\bibfnamefont {P.}~\bibnamefont
  {Demetriou}}, \bibinfo {author} {\bibfnamefont {C.}~\bibnamefont {Grama}}, \
  and\ \bibinfo {author} {\bibfnamefont {S.}~\bibnamefont {Goriely}},\ }\href
  {\doibase 10.1016/s0375-9474(02)00756-x} {\bibfield  {journal} {\bibinfo
  {journal} {Nucl. Phys. A}\ }\textbf {\bibinfo {volume} {707}},\ \bibinfo
  {pages} {253} (\bibinfo {year} {2002})}\BibitemShut {NoStop}%
\bibitem [{\citenamefont {Avrigeanu}\ \emph {et~al.}(2014)\citenamefont
  {Avrigeanu}, \citenamefont {Avrigeanu},\ and\ \citenamefont
  {M{\u{a}}n{\u{a}}ilescu}}]{Avrigeanu14-PRC}%
  \BibitemOpen
  \bibfield  {author} {\bibinfo {author} {\bibfnamefont {V.}~\bibnamefont
  {Avrigeanu}}, \bibinfo {author} {\bibfnamefont {M.}~\bibnamefont
  {Avrigeanu}}, \ and\ \bibinfo {author} {\bibfnamefont {C.}~\bibnamefont
  {M{\u{a}}n{\u{a}}ilescu}},\ }\href {\doibase 10.1103/PhysRevC.90.044612}
  {\bibfield  {journal} {\bibinfo  {journal} {Phys. Rev. C}\ }\textbf {\bibinfo
  {volume} {90}},\ \bibinfo {pages} {044612} (\bibinfo {year} {2014})},\
  \Eprint {http://arxiv.org/abs/1406.1656} {1406.1656} \BibitemShut {NoStop}%
\bibitem [{\citenamefont {Nolte}\ \emph {et~al.}(1987)\citenamefont {Nolte},
  \citenamefont {Machner},\ and\ \citenamefont {Bojowald}}]{Nolte87-PRC}%
  \BibitemOpen
  \bibfield  {author} {\bibinfo {author} {\bibfnamefont {M.}~\bibnamefont
  {Nolte}}, \bibinfo {author} {\bibfnamefont {H.}~\bibnamefont {Machner}}, \
  and\ \bibinfo {author} {\bibfnamefont {J.}~\bibnamefont {Bojowald}},\ }\href
  {\doibase 10.1103/PhysRevC.36.1312} {\bibfield  {journal} {\bibinfo
  {journal} {Phys. Rev. C}\ }\textbf {\bibinfo {volume} {36}},\ \bibinfo
  {pages} {1312} (\bibinfo {year} {1987})}\BibitemShut {NoStop}%
\bibitem [{\citenamefont {Avrigeanu}\ \emph {et~al.}(1994)\citenamefont
  {Avrigeanu}, \citenamefont {Hodgson},\ and\ \citenamefont
  {Avrigeanu}}]{Avrigeanu94-PRC}%
  \BibitemOpen
  \bibfield  {author} {\bibinfo {author} {\bibfnamefont {V.}~\bibnamefont
  {Avrigeanu}}, \bibinfo {author} {\bibfnamefont {P.~E.}\ \bibnamefont
  {Hodgson}}, \ and\ \bibinfo {author} {\bibfnamefont {M.}~\bibnamefont
  {Avrigeanu}},\ }\href {\doibase 10.1103/PhysRevC.49.2136} {\bibfield
  {journal} {\bibinfo  {journal} {Phys. Rev. C}\ }\textbf {\bibinfo {volume}
  {49}},\ \bibinfo {pages} {2136} (\bibinfo {year} {1994})}\BibitemShut
  {NoStop}%
\bibitem [{\citenamefont {Mohr}\ \emph {et~al.}(2013)\citenamefont {Mohr},
  \citenamefont {Kiss}, \citenamefont {F\"ul\"op}, \citenamefont {Galaviz},
  \citenamefont {Gy\"urky},\ and\ \citenamefont {Somorjai}}]{Mohr13-ADNDT}%
  \BibitemOpen
  \bibfield  {author} {\bibinfo {author} {\bibfnamefont {P.}~\bibnamefont
  {Mohr}}, \bibinfo {author} {\bibfnamefont {G.}~\bibnamefont {Kiss}}, \bibinfo
  {author} {\bibfnamefont {Z.}~\bibnamefont {F\"ul\"op}}, \bibinfo {author}
  {\bibfnamefont {D.}~\bibnamefont {Galaviz}}, \bibinfo {author} {\bibfnamefont
  {G.}~\bibnamefont {Gy\"urky}}, \ and\ \bibinfo {author} {\bibfnamefont
  {E.}~\bibnamefont {Somorjai}},\ }\href {\doibase 10.1016/j.adt.2012.10.003}
  {\bibfield  {journal} {\bibinfo  {journal} {At. Data Nucl. Data Tables}\
  }\textbf {\bibinfo {volume} {99}},\ \bibinfo {pages} {651} (\bibinfo {year}
  {2013})},\ \Eprint {http://arxiv.org/abs/1212.2891} {1212.2891} \BibitemShut
  {NoStop}%
\bibitem [{\citenamefont {Mohr}(2019)}]{Mohr19-IJMPE}%
  \BibitemOpen
  \bibfield  {author} {\bibinfo {author} {\bibfnamefont {P.}~\bibnamefont
  {Mohr}},\ }\href {\doibase 10.1142/S0218301319500290} {\bibfield  {journal}
  {\bibinfo  {journal} {Int. J. of Mod. Phys. E}\ }\textbf {\bibinfo {volume}
  {28}},\ \bibinfo {pages} {1950029} (\bibinfo {year} {2019})},\ \Eprint
  {http://arxiv.org/abs/1905.02634} {1905.02634} \BibitemShut {NoStop}%
\bibitem [{\citenamefont {Brown}\ and\ \citenamefont
  {Rho}(1981)}]{Brown81-NPA}%
  \BibitemOpen
  \bibfield  {author} {\bibinfo {author} {\bibfnamefont {G.}~\bibnamefont
  {Brown}}\ and\ \bibinfo {author} {\bibfnamefont {M.}~\bibnamefont {Rho}},\
  }\href {\doibase 10.1016/0375-9474(81)90043-9} {\bibfield  {journal}
  {\bibinfo  {journal} {Nucl. Phys. A}\ }\textbf {\bibinfo {volume} {372}},\
  \bibinfo {pages} {397} (\bibinfo {year} {1981})}\BibitemShut {NoStop}%
\bibitem [{\citenamefont {Mohr}(2000)}]{Mohr00-PRC}%
  \BibitemOpen
  \bibfield  {author} {\bibinfo {author} {\bibfnamefont {P.}~\bibnamefont
  {Mohr}},\ }\href {\doibase 10.1103/PhysRevC.61.045802} {\bibfield  {journal}
  {\bibinfo  {journal} {Phys. Rev. C}\ }\textbf {\bibinfo {volume} {61}},\
  \bibinfo {pages} {045802} (\bibinfo {year} {2000})},\ \Eprint
  {http://arxiv.org/abs/nucl-th/0001059} {nucl-th/0001059} \BibitemShut
  {NoStop}%
\bibitem [{\citenamefont {Rauscher}(2010)}]{Rauscher10-PRC}%
  \BibitemOpen
  \bibfield  {author} {\bibinfo {author} {\bibfnamefont {T.}~\bibnamefont
  {Rauscher}},\ }\href {\doibase 10.1103/PhysRevC.81.045807} {\bibfield
  {journal} {\bibinfo  {journal} {Phys. Rev. C}\ }\textbf {\bibinfo {volume}
  {81}},\ \bibinfo {pages} {045807} (\bibinfo {year} {2010})},\ \Eprint
  {http://arxiv.org/abs/1003.2802} {1003.2802} \BibitemShut {NoStop}%
\end{thebibliography}%
\end{document}